\documentclass[amsmath,amssymb,amsbsy,twocolumn,prb,showpacs,superscriptaddress]{revtex4-1}

\usepackage{amsfonts} 
\usepackage{amsmath}
\usepackage{amssymb}
\usepackage{cancel} 
\usepackage{bm}
\usepackage[usenames, dvipsnames]{color} 
\usepackage{dcolumn}
\usepackage{epic} 
\usepackage{epsfig}
\usepackage{graphicx}
\usepackage[breaklinks,colorlinks = true,linkcolor = red,urlcolor=cyan,citecolor=blue]{hyperref}
\usepackage{latexsym} 
\usepackage{mathrsfs}
\usepackage{soul}
\usepackage{subfigure}
\usepackage{times}
\usepackage{wrapfig}
\usepackage{wasysym}
\usepackage{xy}
\usepackage{dsfont}

\newcommand\undermat[2]{%
  \makebox[0pt][l]{$\smash{\underbrace{\phantom{%
    \begin{matrix}#2\end{matrix}}}_{\text{$#1$}}}$}#2}

\begin{document}

\title{A classification of magnetic frustration and metamaterials from topology}

\author{Krishanu Roychowdhury}
\affiliation{Laboratory of Atomic And Solid State Physics, Cornell University, Ithaca, NY 14853.}
\affiliation{Kavli Institute for Theoretical Physics, University of California, Santa Barbara, CA 93106-4030.}
\author{Michael J. Lawler}
\affiliation{Laboratory of Atomic And Solid State Physics, Cornell University, Ithaca, NY 14853.}
\affiliation{Kavli Institute for Theoretical Physics, University of California, Santa Barbara, CA 93106-4030.}
\affiliation{Department of Physics, Applied Physic and Astronomy, Binghamton University, Binghamton, NY, 13902.}


\begin{abstract}
We study the relationship between the physics of topology and zero modes in frustrated systems and metamaterials. Zero modes that exist in topological matters are distinct from the ones arising from symmetry breaking. Incidentally, a prominent aspect of frustrated systems and metamaterials also is to harbor such kind of zero modes in form of an {\it accidental} degeneracy. Taking cues from these two apparently different phenomena, we ask a simple question: are the robust features of frustration topologically protected and if so can we classify different types of frustration using topology? In answering these questions we invoke the tools of topological mechanics to identify the key agent at play, namely the rigidity matrix, which is a non-Hermitian matrix and decides the topology of spin-wave zero modes in a frustrated magnet or phonon modes in metamaterials. Further developments of the theory rely on combining the recent developments in our understanding of Maxwell constraint counting and generalizing the ten-fold way classification of Hermitian matrices to non-Hermitian matrices. The result is a three-fold way classification for each Maxwell counting index. We illustrate the classification by demonstrating the existence of a new vortex-like invariant for real rigidity matrices using random matrices and through example frustrated spin models. So by classifying all the rigidity matrices, we answer the question of the origin of frustration (i.e. zero modes in the form of accidental degeneracy) in a wide class of frustrated magnets and metamaterials by linking it to topological invariants.
\end{abstract}


\maketitle
\section{Introduction}

A remarkable aspect of frustrated systems and metamaterials is the vast {\it accidental degeneracy} of ground states~\cite{lacroix2011introduction, diep2013frustrated} which manifests as zero modes that are not associated to any symmetry breaking, in other words, distinct from the Goldstone modes. Notable examples, for instance in the magnetism side, include the magnon flat band in the ideal kagome Heisenberg antiferromagnet (KHAF) for which candidate materials are plentiful. Besides the flat band, there exist other forms of zero modes such as line nodes in anisotropic kagome materials~\cite{roychowdhury2017spin}, nodal points in mechanical lattices~\cite{po2016phonon, rocklin2016mechanical}, and two dimensional surfaces of zero modes in diamond lattice spinels~\cite{bergman2007order}. Surprisingly, these zero modes are immune to classes of perturbations counterintuitively implying a manifold of zero modes signifies a robust nature of frustration.

Consider the distorted kagome antiferromagnet ${\rm Cs}_2{\rm CeCu}_3{\rm F}_{12}$. In Ref.~[\onlinecite{roychowdhury2017spin}], we predicted nodal lines in their spin-wave band structures at experimentally determined exchange interactions. This prediction was not an accident. Assuming spin dynamics in these materials are dominated by nearest neighbor (nn) exchanges we could map their ground states onto exotic spin origami analogs. Building on the modern theory of topological mechanics associated with origami~\cite{chen2016topological}, we were able to show these line nodes came from a change in a $\mathbb{Z}_2$ topological invariant across the Brillouin zone (BZ). The situation is vividly reminiscent to one we encounter in Weyl semimetals with topologically protected bulk zero modes~\cite{rao2016weyl, yan2017topological,armitage2018weyl}. The difference is that more than symmetry is needed to protect the topological invariant. The resemblance encourages the question: are zero modes in frustrated systems demanded by a change in some topology? If so, perhaps a classification of the underlying topology can enable us to explore new varieties of frustration.

The past decade has already witnessed the laudable achievements of a topological classification adding novelties to the simple band theory of electrons and prophesying new states of quantum matter as consequences. In a succinct form, it is the table of the {\it ten-fold way} that captures different topology of band structures in electronic insulators and superconductors and provides an exhaustive list of free fermion topological phases~\cite{zirnbauer1996mr, altland1997nonstandard, kitaev2009periodic, ryu2010topological, ryu2012electromagnetic} (for more relevant references and a review of the ten-fold classification, see Ref.~[\onlinecite{ludwig2015topological}] and the references therein). These phases arise in absence or presence of certain symmetries of the Hamiltonian which lay the cornerstone of the classification problem. 

The ten-fold way has successfully enabled the unveiling of new topological states of matter several of which were elusive prior to the inception of classification; of notable mention are topological superconductors~\cite{schnyder2008classification, kitaev2009periodic, ryu2010topological, ryu2012electromagnetic}. By virtue of this classification, we have now found the existence of five distinct topological insulators/superconductors in every dimension, some of which have also nucleated experimental activities. Spirited with a similar ideology, we attempt to classify the topology of zero modes in frustrated systems and metamaterials and illuminate the origin of frustration in the form of accidental degeneracy by linking it to topological invariants with the hope that this classification will also lead to new experimental activities.   

In two seminal papers~\cite{moessner1998properties, moessner1998low}, Moessner and Chalker presented an elementary understanding of frustration in spin systems using Maxwell counting that can shed some light on the robust nature of the degeneracy. The key idea is to group the terms in the Hamiltonian into constraints following which a naive degeneracy estimate $\nu$ is obtained by having fewer constraints $K$ than the degrees of freedom (d.o.f) $D$, i.e. $\nu = D - K$, which we call the Moessner-Chalker-Maxwell (MCM) index in this paper. The index also caters a perspective on the problem of lifting the degeneracy by perturbations. If the perturbations do not introduce new constraints but only deform them, the degeneracy should persist. But this understanding is incomplete -- it relies on a “naive” estimate that ignores linear dependence among the constraints. So this needs to be taken into account.  For example, the ideal classical KHAF has many zero modes. But its spin Hamiltonian can be written as
\begin{equation}
 H = J\sum_{\langle i,j\rangle} {\bf S}_i\cdot{\bf S}_j = \frac{J}{2} \sum_\triangle {\bf S}_{\triangle}^2 + {\rm const.},
 \label{ham2}
\end{equation}
where ${\bf S}_{\triangle}={\bf S}_{i}+{\bf S}_{j}+{\bf S}_{k}$ is the total spin in the triangle $ijk$ and each unit cell has two such triangles. So there are six constraints per unit cell but also six degrees of freedom (three spin unit vectors) leading to $\nu=0$ per unit cell. But the same model on the pyrochlore lattice leads by the same argument to $\nu = 2$ per unit cell. So it would seem distortions that preserve the number of constraints could lift all the degeneracy of the kagome antiferromagnets by rendering them linearly independent but could not lift all the degeneracy of the pyrochlore lattice. In this way, Moessner-Chalker-Maxwell counting can predict a kind of topologically protected set of zero modes in some systems: if $\nu>0$ a set of zero modes exists so long as perturbations change the form of the constraints and not their number.  

Remarkably, the $\nu=0$ point has enticed much attention following the seminal work of Kane and Lubensky who discovered the possibility of topological protection of zero modes in mechanical systems~\cite{kane2014topological} even in this case. A number of further studies have emerged making further advancements in that field~\cite{rocklin2016mechanical} including the ones involving metamaterials for their exotic properties owing to these same zero modes~\cite{paulose2015selective, paulose2015topological, rocklin2015transformable, abbaszadeh2016sonic}. In essence, these ``isostatic'' systems have both an energy gap and $\nu=0$ for periodic boundary conditions. Then for open boundary conditions $\nu>0$ and a zero mode such as an edge state arises. Kane and Lubensky further related this observation to a topological invariant through a mapping to a fermion-like band structure. However, these results at $\nu=0$ seem like a special case and it is not obvious whether there is any extension to $\nu>0$ or $\nu<0$. But there are many frustrated magnets in this latter group both since it has been a goal of the field to find magnets which are as highly frustrated as possible ($\nu>0$) and because in the search for such magnets, many are found which are less frustrated ($\nu<0$) but still show signs of frustration such as those whose frustration arises not from the geometry of the underlying lattice but as a consequence of competing exchange interactions between the spins. So while we search for topological protection of frustration in special $\nu=0$ antiferromagnets, if similar underlying ideas imply topological protection of frustration in the vast majority of frustrated magnets, we are bound to understand the experimental significance of frustration from topology as it applies to solid state systems.   

Here we solve the problem of identifying topology in zero modes of mechanical systems and frustrated magnets at any $\nu$ including $\nu \neq 0$. We do so by classifying the rigidity matrix ($\mathcal{R}$) that characterizes the constraints. In the example of the ideal kagome Heisenberg model above, this matrix is related to the nonlinear constraint functions ${\bf S}_{\triangle}$ simply by linearizing them in a spin-wave expansion
\begin{equation}
S_{\triangle\alpha} = \mathcal{R}_{\triangle\alpha,i\mu} x^{i\mu} ~~;~~ \mathcal{R}_{\triangle\alpha,i\mu} = \frac{\partial S_{\triangle\alpha}}{\partial x^{i\mu}},
\label{riggmat}
\end{equation}
were $x^{i\mu}$ are the spin-wave coordinates with $i$ labeling the sites, $\mu$ the polar or azimuthal components, and $\mathcal{R}$ is the non-Hermitian rigidity matrix. The spin-wave coordinates $x^{i\mu}$ consist of pairs of canonically conjugate d.o.fs $q^i,p_i$ with $\{q^i,p_i\}=1$ which specify a spin of unit magnitude as ${\bf S}_i=(\cos(q^i)\sqrt{1-p_i^2},\sin(q^i)\sqrt{1-p_i^2},p_i)$ retaining the spin algebra $\{S_{i\alpha},S_{i\beta}\}=\epsilon^\gamma_{\alpha\beta} S_{i\gamma}$. Any change in the Hamiltonian which preserves the number of constraints therefore just deforms $\mathcal{R}$. But any spin-wave $x^{i\mu}$ which lives in the null space of this matrix is a zero mode. So a classification of these matrices directly addresses the question of how frustration could be preserved by perturbations. We show such a classification can indeed be constructed by extending some of the methods used to construct the ten-fold way classification of electronic systems from Hermitian matrices to non-Hermitian matrices. The results maps rigidity matrices onto either classical Lie groups or Stiefel manifolds which are well studied topological spaces whose homotopy groups are all worked out in the mathematics literature~\cite{saito1955homotopy, james1958intrinsic, matsunaga1959homotopy, gilmore1967complex, mori1972homotopy, james1976topology, dodson1997user}. They are also reminiscent of the topology discovered recently in self energies also viewed as non-Hermitian matrices\cite{zhou2018observation}.These homotopy spaces reproduce the Kane and Lubensky $\nu=0$ topological invariants but also show there are plenty of other such invariants both for $\nu=0$ and $\nu\neq 0$. We then demonstrate this latter discovery by taking this mathematics and apply it to several examples, including even the $J_1-J_2$ square lattice antiferromagnet. These examples suggest non-trivial topological invariants exist in essentially all models of frustrated magnetism, that zero modes related to this frustration are likely a manifestation of when the topological invariant changes and perhaps most importantly, that there are perturbations which would preserve their frustration broadening our search for exotic phases of matter in frustrated magnets. 

The paper is organized as follows. In the following section (Sect.~\ref{sectwo}), we discuss the scopes of exploring the topology by means of rigidity matrices in metamaterials/frustrated systems whose classification constitutes the theme of the present work. In Sect.~\ref{secthree}, we sketch the concepts of the previously known symmetry-based classification of Hamiltonian matrices and mention its limitations in the study of the topology of frustration. This brings us to introduce our new scheme of obtaining an appropriate classifying space of rigidity matrices whose topology is the sought for object to explain the origin of frustration. We present the mathematical details in Sect.~\ref{secfour}. The topology of this space is explored in details (in Sect.~\ref{secfive}) and discussed in the form of classification tables using the examples of random matrices for both $\nu=0$ and $\nu\neq0$ systems. In Sect.~\ref{secsix}, we illustrate our classification tables further by exemplifying a variety of frustrated magnets in which the zero modes are demanded from topology. We also depict the explicit constructions of different topological invariants for those models before we finally conclude in Sect.~\ref{secseven} summarizing the important results on the classification of magnetic frustration from topology.        

\section{Fine-tuning from solid state physics to metamaterials}\label{sectwo}

We define frustration as an accidental degeneracy arising from fine tuning. Largely, the study of frustrated systems has focused on a limited set of models, but, many systems, such as the distorted kagome antiferromagnets and diamond lattice spinels mentioned above, exhibit frustration over a range of parameters in the model. As such, frustration is both delicate and robust. Here we are proposing a classification of frustrated systems. For concreteness, let us focus on quadratic spin Hamiltonians common in solid state systems that can be written in the form
 \begin{equation}
  H_{\rm spin} = \frac{1}{2} S_{i\alpha} J^{i\alpha;j\beta} S_{j\beta}, 
  \label{Hspingen}
 \end{equation}
where $J^{i\alpha,j\beta}$ is an exchange matrix and $\alpha\in\{x,y,z\}$ denotes the spin components. It is known~\cite{moessner1998properties, moessner1998low, lawler2016supersymmetry}, that $H_{\rm spin}$ can be recognized as a sum of positive definite terms such as $(S_{i\alpha} + S_{j\beta})^2$ on bond $ij$ if we reorganize terms in the Hamiltonian and add a constant shift to the energy. In systems with short-range interactions, these terms can act like constraints on a set of spins which are localized over small clusters of the lattice when they are satisfied individually. In this case, they are ``frustration-free'' in the technical sense that all terms when organized in this fashion are satisfied in the ground state. For example, the ideal triangular Heisenberg antiferromagnet can be understood as imposing a vanishing total spin on each triangle. A simultaneous minimization of all the constraints, however, can still lead to an accidental degeneracy of zero modes (i.e. an underconstrained system). For some systems, this degeneracy is even extensive growing with the system size. But it is hard to classify systems by these constraints because it is not yet established when or how terms in Hamiltonians can be reorganized into constraints outside of examples from model systems. 
 
Remarkably, mechanical metamaterials could offer clues to the classification problem. They are also among the systems that are exclusively known to foster such a vast degeneracy of zero modes. This is because the underlying Hamiltonian also contains frustration-free constraint functions as mentioned above which often leave them underconstrained. They are engineered to have the degeneracy. Since this is really a defining characteristic of a metamaterial, we call the class of Hamiltonians with the specific form 
 \begin{equation}
 H_{\rm meta} = \frac{1}{2} L_{m} K^{mn} L_{n}, 
 \label{Hmetamat}
 \end{equation}
``metamaterial Hamiltonians'', where $L_m$ denotes a constraint which could be, for example, an extension of a spring, a momenta that should vanish at zero energy or other constraints on the degrees of freedom. The matrix $K$ should be positive definite to ensure $L_m=0$ in the ground state. As a result, some solid state systems, those whose Hamiltonians can be reorganized into constraints, are naturally occurring metamaterials. 

\begin{figure}
\centering    
\includegraphics[width=66mm]{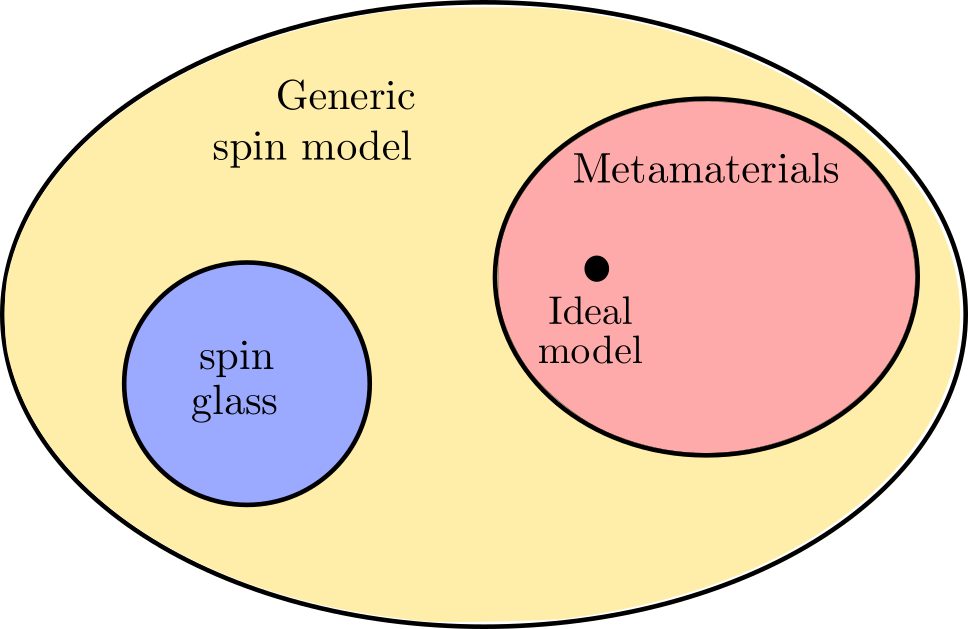}
\caption{A conceptual picture of the space of classical spin models projected onto a two-dimensional space of parameters. The class of metamaterial Hamiltonians defined in Eq.~\ref{Hmetamat} is a fine-tuned case of the generic spin model in Eq.~\ref{Hspingen}. The ideal model (with isotropic nn exchanges as in Eq.~\ref{ham2}) stands as an isolated point in the space of metamaterial Hamiltonians. In the projected picture, perturbations along the perpendicular directions drive a system away from the concerned space. A model of a spin glass which cannot be written as a set of constraints  on the ground state (i.e. the form $H_{\rm meta}$) lies outside the space of metamaterial Hamiltonians.}
\label{fig:hamsp}
\end{figure}

Let us then return to frustration in a solid state system and how it is a delicate phenomenon involving fine tuning. In terms of the above discussion, this fine tuning is in the sense that rearranging the terms of the spin Hamiltonian $H_{\rm spin}$ such that it acquires a form like $H_{\rm meta}$ (with $L_m$ as functions of $S_{i\alpha}$) limits $H_{\rm meta}$ to a subset of $H_{\rm spin}$ (Fig.~\ref{fig:hamsp}). For example, such a fine-tuning enables one to express  the nn spin model in an ideal KHAF in terms of the constraint functions given by ${\bf S}_{\triangle}$ (Eq.~\ref{ham2}). But, provided couplings on each triangle obey a certain triangle inequality condition, we can also write the entire space of nearest neighbor KHAFs in this form\cite{roychowdhury2017spin}. So the engineering associated with metamaterials suggests there is a space of Hamiltonians all sharing similar characteristics and so define a class of frustration and that these classes arise naturally in solid state physics by the locality of interactions in insulators.

The simplest possible model of a mechanical metamaterial can be regarded as a collection of coupled oscillators. The normal modes of the oscillation are obtained by solving the equation of motion ${\ddot{{\bf x}}}=-D\cdot{\bf x}$, where ${\bf x}$ lists the displacements of the mass points. The matrix $D$ is known as the ``dynamical matrix'' whose eigenvalues, when square rooted and scaled appropriately, yield the normal mode frequencies. An example could be the system of classical phonons represented by vibrational modes of balls connected by Hookean springs. Assuming the balls of unit mass and the springs having unit spring constant, the Hamiltonian of the system is
 \begin{equation}
 H_B = \frac{1}{2} \bigg(\sum_j p_j^2+ \sum_m e_m^2 \bigg),
 \label{springham1}
 \end{equation}
 where $p_j$ is the momentum of the $j$-th ball and $e_m$ is the extension of the $m$-th spring. The spring extensions are related to the displacements of the balls from their equilibrium positions as 
 \begin{equation}
 {\bf e}=A\cdot\bf{x}+\mathcal{O}({{\bf x}}^2).
 \label{rig4}
 \end{equation}
 The matrix $A$ is known as the ``compatibility matrix'' ($A^T$ is known as the ``equilibrium matrix'')~\cite{kane2014topological}. In a harmonic approximation, only the leading linear term contributes yielding 
 \begin{equation}
 H_B = \frac{1}{2} \bigg[{\bf p}^2+ (A\cdot\bf{x})^2 \bigg],
 \label{springham2}
 \end{equation}
 where ${\bf p}$ is the column vector consisting of the momenta of the mass points. The Lagrangian equation of motion for ${\bf x}$ then implies $D=A^TA$, in other words, $A$ is the square root matrix of $D$. For the model system described above, the d.o.fs are ${\bf p}$ and ${\bf x}$ while the constraints are that in the ground state, ${\bf p}=0$ and $A\cdot{\bf x}=0$~\cite{lawler2016supersymmetry}. So the rigidity matrix ($\mathcal{R}$) defined in Eq.~\ref{riggmat} will be of the form $\mathcal{R}={\rm diag}(1, A)$.

Let us now view the situation in a frustrated magnet. A fine-tuning depending on the structure of the underlying lattice and the ranges of the interactions may enable us to write $H_{\rm spin}$ in Eq.~\ref{Hspingen} as
 \begin{equation}
  H_{\rm spin} = \frac{1}{2} S_{\triangle\alpha} J^{\triangle\alpha,\triangle'\beta} S_{\triangle'\beta}, 
  \label{Hspinfrust}
 \end{equation}
where $S_{\triangle\alpha}$ is the constraint on a simplex $\triangle$ of the lattice involving the spin component $\alpha$. If $J^{\triangle\alpha,\triangle'\beta}=J\delta^{\triangle\triangle'}\delta^{\alpha\beta}$, the model represents nn Heisenberg model on a geometrically frustrated lattice. For a triangular simplex, it is the KHAF; for a crisscross square simplex, it is the checkerboard lattice Heisenberg model; for a tetrahedral simplex, it is the pyrochlore magnet as first identified in Ref.~[\onlinecite{moessner1998properties}] and [\onlinecite{moessner1998low}]. For the sake of categorization, we term the models with this particular type of Hamiltonian, the ``ideal models'' which is specified by a diagonal $J$ matrix proportional to identity, so represents an isolated point in the space of metamaterial Hamiltonians (Fig.~\ref{fig:hamsp}).     
 
Unfrustrated spin models can also have Hamiltonian expressed as a quadratic function of constraints as in Eq.~\ref{Hspinfrust}. A simple example could be the case of nn Heisenberg model on a square lattice with the Hamiltonian
 \begin{equation}
 H = J\sum_{\langle i,j\rangle} {\bf S}_i\cdot{\bf S}_j \equiv \frac{1}{2} L_{m\alpha} J^{m\alpha,n\beta} L_{n\beta},
 \label{ham1}
 \end{equation}
with $J^{m\alpha,n\beta}=J\delta^{mn}\delta^{\alpha\beta}$; $m$ labels the nn bonds $\langle i,j\rangle$ and $\alpha$ the spin vector components. Evidently the constraints for the ground state are $L_{m\alpha} \equiv S_{i\alpha} + {\rm sgn}(J) S_{(i+\hat{x})\alpha} = 0$ on each horizontal bond and $L_{m\alpha} \equiv S_{i\alpha} + {\rm sgn}(J) S_{(i+\hat{y})\alpha} = 0$ on each vertical bond. Up to global spin rotations, this uniquely selects the ground state to be the uniform state (N\'{e}el state) for $J<0$ ($J>0$). Later we will include other spin systems in which the Hamiltonian can be expressed to have a similar form to Eq.~\ref{Hmetamat}. They all represent magnetic analogs of mechanical metamaterials to which one can apply the theory of rigidity matrix to analyze the (linearized) zero modes. As argued before, these zero modes reside in the null space of the rigidity matrix. It also provides important clues to unfurl the topological aspects of the zero modes over a broad range of solid state systems from spin-waves in microscopic spin models~\cite{lawler2016supersymmetry, roychowdhury2017spin} to phonons in macroscopic metamaterial systems~\cite{po2016phonon, rocklin2016mechanical}. 

The MCM index $\nu$ in terms of the rigidity matrix $\mathcal{R}$ reads~\cite{kane2014topological, lawler2016supersymmetry}  
 \begin{equation}
  \begin{split} 
   \nu &= {\rm Cols}[\mathcal{R}] - {\rm Rows}[\mathcal{R}] \\
       &= {\rm Rank}[\mathcal{R}] + {\rm null}[\mathcal{R}] - {\rm Rank}[\mathcal{R}^T] - {\rm null}[\mathcal{R}^T] \\
       &= {\rm null}[\mathcal{R}] - {\rm null}[\mathcal{R}^T] \\
       &= N_0 - N_s,
  \end{split} 
  \label{nurig}
 \end{equation}
where we have used ${\rm Rank}[\mathcal{R}]={\rm Rank}[\mathcal{R}^T]$ by the fundamental theorem of linear algebra. From the definition in Eq.~\ref{nurig} it is evident that the index $\nu$ remains invariant as long as the dimensions of $\mathcal{R}$ are unaltered. Although the inputs to calculate $\nu$ have information about the topology of the lattice, it should not be regarded as a true topological invariant that could differentiate between various forms of zero modes such as nodal points, lines, or surfaces or could characterize them. It certainly provides an estimate of them but is incapable of revealing any insight into their topological nature. These properties rely on the structure of the space of $\mathcal{R}$, and not merely its shape. Our study emphasizes the fact that the topology of frustration is intimately linked to that of the classifying space of $\mathcal{R}$ -- to appreciate the former, the latter is the key agent to inspect.  

Symmetries play a central role to classify random (Hermitian) matrices as argued in previous studies concerning disordered fermionic~\cite{altland1997nonstandard} and bosonic systems~\cite{gurarie2003bosonic}. Different symmetry classes have distinct implications on the energy level statistics of a fermionic Hamiltonian~\cite{dyson1962threefold} and the observables derived from that, or scaling of the density of low-energy excitations in a disordered bosonic medium~\cite{gurarie2003bosonic}. This motivates us to develop a symmetry based classification of rigidity matrices which explains the manifestation of the accidental degeneracy of ground states in different forms of zero modes in different frustrated systems. The remaining task is to identify the topology that characterizes these distinct forms. This is what the ten-fold way has achieved in the electronic problems. As our method bears similarities to the homotopy group analysis that lead to the compact table of the ten-fold way, reviewing some of the important mathematical concepts of that classification would help us set the stage in the following section. We will then illuminate the topological aspects of non-Hermitian matrices by considering ensembles of random rigidity matrices under symmetry imposition which determines the topology of the zero modes associated with them. 

\section{A review of the ten-fold way}\label{secthree} 

Classification and characterization of phases have always been a major theme of solid state research. Initiated with Dyson's pioneering work on the three-fold classification of random matrices~\cite{dyson1962threefold}, recent developments have now achieved a complete ten-fold classification of disordered fermionic Hamiltonians~\cite{zirnbauer1996mr, altland1997nonstandard}. The ten distinct classes, named Altland-Zirnbauer (AZ) classes, belong to Cartan's list of classical compact symmetric spaces which characterize the time-evolution operator $e^{iHt}$ for a random Hamiltonian $H$. An extension of Altland-Zirnbauer classification (based on a set of discrete symmetries) to random nonHermitian matrices was also attempted~\cite{bernard2002classification}, however, the topological aspects of those matrices were not illuminated.

Shortly after those developments, relevance of symmetry classes to understand the generic aspects of a random bosonic model was emphasized in a study by Gurarie and Chalker~\cite{gurarie2003bosonic}. Drawing parallels to specific AZ classes in the electronic problems, the authors could evince discerning properties of the low-energy excitations of a random quadratic bosonic Hamiltonian 
\begin{equation}
 H_B=\mathcal{R}^T\mathcal{R}.
 \label{Hboson}
\end{equation}
The excitation frequencies ($\omega$), eigenvalues of the non-Hermitian equations of motion matrix $\sigma_2 H_B$, as they argued, can be derived from an auxiliary fermionic problem with a chiral (Hermitian) Hamiltonian matrix
\begin{equation}
 H_F = \mathcal{R}\sigma_2\mathcal{R}^T ~~;~~ \sigma_2=
 \begin{pmatrix}
  0 & -i \\
  i & 0
 \end{pmatrix}.
 \label{Hfermion}
\end{equation}
This foreshadowed the supersymmetry associated with a system described by a rigidity matrix exploited in recent studies\cite{kane2014topological, lawler2016supersymmetry}. For a range of bosonic models, $\mathcal{R}$ can be identified in a continuous or discrete form. The spectral properties of $H_F$~\cite{beenakker2015random} can then be exploited directly to observe distinct features in the scaling of the density of excitations $\rho(\omega)$ of both Goldstone and non-Goldstone types.   

A few years later, at the wake of topological insulators spreading huge excitements in the community, a complete classification of gapped and gapless topological matters was brought into existence exploiting the homotopy theory of topological equivalence~\cite{kitaev2009periodic, schnyder2008classification, ryu2010topological}. Remarkably, the topology is found to have ten distinct classes as discovered earlier by Altland and Zirnbauer, and can be compactly described in a periodic table~\cite{kitaev2009periodic}, popularly called the ``ten-fold way''. 

The ten-fold way classification of electronic band structures defines two Hamiltonians as equivalent if they can be smoothly deformed into each other without closing the energy gap between occupied and unoccupied bands. A violation of this topology then demands the gap must close. One way to understand this is to flatten the bands which is well defined only if the bands are separated and violations relate directly to the gap closing. Specifically, without closing the excitation gap, one can smoothly deform $H({\bf k})$, the Bloch Hamiltonian, to the spectrally flattened Hamiltonian $\tilde{H}({\bf k})={\rm sign}[H({\bf k})]$~\cite{kitaev2006anyons, ludwig2015topological} defined as
\begin{equation}
 \tilde{H}({\bf k}) = \mathcal{U}({\bf k}) \begin{pmatrix}
                                            \mathds{1}_M  & 0 \\
					    0 & -\mathds{1}_N 
                                           \end{pmatrix} \mathcal{U}^\dagger({\bf k}),
\label{ham3}
\end{equation}
with $M$($N$) bands above(below) the chemical potential. The space of these matrices, which depends on the symmetries of  $\tilde{H}({\bf k})$, is then a manifold which is either topologically trivial or non-trivial. If the manifold has a non-trivial topology, it then remains to determine how this topology can be broken and where the energy gap closes as a consequence. 

\begin{figure}
\centering    
\includegraphics[width=1.0\columnwidth]{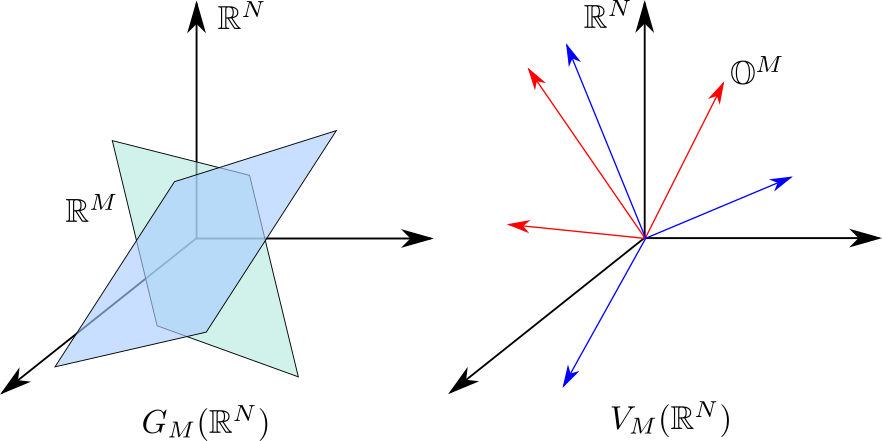}
\caption{Left: A conceptual picture of the real Grassmannian manifold $G_M(\mathds{R}^{N})$ which is a set of all $M$-plane in $\mathds{R}^{N}$ (two of them are shown). Right: The same for the real Stiefel manifold $V_M(\mathds{R}^{N})$ which is a set of all $M$-frame in $\mathds{R}^{N}$ (two of them are shown). To each point in $G_M(\mathds{R}^{N})$, i.e. a specific $M$-plane, there corresponds in $V_M(\mathds{R}^{N})$ the set of all orthogonal bases (formed by $M$ orthonormal real vectors) for that plane.}
\label{fig:manifold}
\end{figure}

In Ref.~[\onlinecite{ryu2010topological}], it was argued that the only relevant symmetries to classify a quantum mechanical Hamiltonian include two antiunitary symmetries: time-reversal ($\mathcal{T}$) and charge conjugation or particle-hole transformation ($\mathcal{C}$) and an anticommuting unitary symmetry: chiral or sublattice symmetry ($\mathcal{S}=\mathcal{T}\cdot\mathcal{C}$). In the absence of any symmetry, the space of $\tilde{H}({\bf k})$ is in one-to-one correspondence with the set of all $N$ dimensional sub-Hilbert spaces (each spanned by the occupied states) arising in the full $M+N$ dimensional Hilbert space.  This is a complex Grassmannian manifold $G_M(\mathds{C}^{M+N})$ (its real analog is shown in Fig.~\ref{fig:manifold}) whose homotopy maps $\pi_d$ in a given dimension $d$ reveals the topology of $H({\bf k})$. These are all known since the complex Grassmannian is equivalent to the coset space $U(M+N)/[U(M)\times U(N)]$. Upon applying the symmetries individually or in different combinations to $H({\bf k})$, the structure of the Grassmannian changes as $\tilde{H}({\bf k})$ acquires additional structures under those operations leading to a ten-fold classification of the topology. In other words, there exist ten topologically distinct classes of Hamiltonians that completely characterize the (noninteracting) electronic systems~\cite{ludwig2015topological}.  

Symmetry based topological classification is now a mature subject, thanks to the extensive work that have addressed the topic using a variety of methods pertinent to different quantum mechanical systems (both interacting and noninteracting). Examples include symmetry protected topological phases~\cite{lu2012theory, vishwanath2013physics, chen2013symmetry, wang2014interacting, gu2014symmetry, kapustin2014symmetry, kapustin2014bosonic}, Kitaev's Majorana models~\cite{o2016classification}, spin systems, especially various models of spin liquids~\cite{wen2002quantum, chen2011, reuther2014classification, bieri2016projective, huang2018classification}. Furthermore, of note for its relation to the present work, Ref.~[\onlinecite{susstrunk2016classification}] was able to study the topology of a wide spectrum of classical mechanical models by transforming the eigenvalue problem to a Hermitian matrix (interpreted as a Bloch Hamiltonian) that retains the structure of the phonon eigenvectors but expressed in terms of auxiliary variables. They could then predict topological features at finite frequencies where a gap between lower frequency bands and upper frequency bands collapses.  However, despite this maturity, the fundamental building block in all cases is a space of Hermitian matrices which is not of direct use to characterize the zero modes in metamaterials and frustrated magnets. The reason is that frustration is nothing but an accidental degeneracy (of ground states) which can not be attributed to the symmetries of a Hamiltonian. It, therefore, is important to seek additional structures beyond a Hermitian operator, a gap between two sets of its eigenvalues and the symmetries. 

\section{Classification of metamaterials}\label{secfour}

In order to understand if zero modes in a metamaterial are stabilized by topology, we explore the space of $\mathcal{R}$ associated with the Hamiltonian $H_{\rm meta}$ by including distortions to $\mathcal{R}$ without changing its dimension, i.e. $\nu$, and ask what is the topology of this space in presence of the symmetries of the problem. 

We will consider the most general form of the rigidity matrix $\mathcal{R}$ obtained from expanding about a specific configuration of a metamaterial system as in Eq.~\ref{riggmat} as a complex matrix. Thus we will view classical systems with no symmetry and real $\mathcal{R}$ as a higher symmetry case of a more general system described by a complex matrix. The complex form is also  a useful starting point for the case when we write $\mathcal{R}$ in a momentum space basis (obtained from the Fourier transformation of the real space basis). 

Now, with a complex non-Hermitian matrix as a starting point, the eigenvalues of $\mathcal{R}$ in general are complex numbers and the spectral flattening technique used in the ten-fold way is inapplicable. We need an alternative way to encode the gap condition of a topological space. 

Remarkably, a simple solution is to spectrally flatten the singular values of $\mathcal{R}$, which are always real and nonnegative. This is a gap condition for systems described by a rigidity matrix because the number of non-zero singular values is the rank of the matrix and the only way a zero mode can be introduced and a gap closed is to reduce this rank. With this gapping condition in hand, we can proceed with the homotopy analysis of the space formed by the resultant flattened singular value matrices containing the singular vectors. 

\subsection{Three-fold way for $\nu=0$ systems}

For a generic $\nu=0$ system, the rigidity matrix is a complex square $M\times M$ matrix (from Eq.~\ref{nurig}). Its singular value decomposition (SVD) then reads
\begin{equation}
 \mathcal{R}=\mathcal{U}\Lambda_R \mathcal{V}^{\dagger},
 \label{svd1}
\end{equation}
where $\mathcal{U},\mathcal{V}\in U(M)$ and $\Lambda_R$ is a diagonal matrix containing $M$ singular values which are all positive provided $\mathcal{R}$ is of full rank. The singular vectors contained in $\mathcal{V}$ are the eigenvectors of $\mathcal{R}^\dagger \mathcal{R}$, the singular vectors contained in $\mathcal{U}$ are the eigenvectors of $\mathcal{R}\mathcal{R}^\dagger$ and the singular values are the square roots of the eigenvalues of either $\mathcal{R}^\dagger \mathcal{R}$ or $\mathcal{R}\mathcal{R}^\dagger$. Like in the ten-fold classification of electronic systems where eigenvalues are flattened, flattening the singular values: $\Lambda_R\rightarrow \tilde{\Lambda}_R=\mathbb{I}_{M\times M}$ produces a new matrix 
\begin{equation}
 Q=\mathcal{U}\tilde{\Lambda}_R\mathcal{V}^{\dagger}=\mathcal{U}\mathcal{V}^{\dagger} 
 \label{Qunit}
\end{equation}
analogous to $\tilde{H}({\bf k})$ in Eq.~\ref{ham3}. We call $Q$ to be the SVD flattened matrix of $\mathcal{R}$. Since $\mathcal{U}$ and $\mathcal{V}$ are unitary matrices, so is $Q$. 
It implies the bosonic and fermionic Hamiltonians $H_B=\mathcal{R}^\dagger\mathcal{R}$ (in Eq.~\ref{Hboson}) and $H_F=\mathcal{R}\sigma_2\mathcal{R}^\dagger$ are diagonalizable by unitary matrices and the topology of $\mathcal{R}$ can be classified by studying the homotopy groups of unitary matrices.

The above is the case when $\mathcal{R}$ has no particular symmetry. Enforcing symmetries on $\mathcal{R}$ alters the structure of $Q$. The simplest case is a ``commuting'' unitary symmetry with transformation law
\begin{equation}
\mathcal{U}: \mathcal{R} \to \mathcal{U}_F\mathcal{R}\mathcal{U}_B^\dagger,
\end{equation}
where the $U_F$ matrix describes the action of this symmetry on the fermion Hamiltonian $H_F$ and the $U_B$ matrix describes its action on $H_B$. As in the case of a Hermitian matrix, the eigenbasis of this symmetry block diagonalizes $H_B$, $H_F$, $\mathcal{R}$ and $Q$. As far as classification is concerned, then, we can follow the ten-fold way example by assuming we work in the block diagonal basis and focus just on a sub-block of $\mathcal{R}$. So while a unitary symmetry changes the structure of $Q$, it doesn't change its class. 

It then remains to consider antiunitary symmetries (like the time-reversal symmetry and particle-hole symmetry in the ten-fold way) and unitary symmetries which ``anticommute'' (like the chiral symmetry in the ten-fold way) with $\mathcal{R}$. Consider first time-reversal symmetry ($\mathcal{T}$) with $\mathcal{T}^2=1$. Mathematically, it acts on $\mathcal{R}$ like
\begin{equation}
\mathcal{T}: \mathcal{R} \to \mathcal{T}_F\mathcal{R}\mathcal{T}_B^{-1},
\label{Tsymmetry}
\end{equation}
where $\mathcal{T}_F = U_{F} K$, $\mathcal{T}_B = U_B K$, $U_F$ and $U_B$ are unitary, $U_FU_F^* = I$, $U_BU_B^* = I$ and $K$ is complex conjugation. $T_F$ is then time reversal for the $H_F$ problem and $T_B$ is time reversal for the $H_B$ problem. Following Dyson~\cite{dyson1962threefold}, we can then find a basis where $U_F$ and $U_B$ are identity matrices and so time-reversal demands that $\mathcal{R}$ is real instead of complex. Thus $H_B$ ($H_F$) is a real symmetric (imaginary antisymmetric) matrix diagonalizable by orthogonal matrices, in other words, {\it orthogonally similar} to a diagonal matrix. Also, since a real matrix has the same singular value decomposition as Eq.~\ref{svd1} but with the unitary matrices $\mathcal{U}$ and $\mathcal{V}$ replaced with orthogonal matrices, 
\begin{equation}
 Q=\mathcal{U}\mathcal{V}^T  
 \label{Qorth}
\end{equation}
is also an orthogonal matrix which has distinct homotopy maps compared to the complex case. 

We can also realize the case with $\mathcal{T}^2=-1$. Here $U_B$ and $U_F$ satisfy $U_BU_B^* = -I$ and $U_FU_F^*=-I$. We can then choose to work in the standard representation where we can write $U_B=\iota\sigma_2$ and $U_F = \iota\sigma_2$. To understand the structure of $\mathcal{R}$ under the action of this $\mathcal{T}$ symmetry, let us introduce the following notation
\begin{equation}
 \mathcal{R}^\# \equiv (\iota\sigma_2) \mathcal{R}^* (\iota\sigma_2)^{-1}~~;~~\mathcal{R}^\$ \equiv (\iota\sigma_2) \mathcal{R}^T (\iota\sigma_2)^{-1}. 
 \label{defs1}
\end{equation}
so that $(\mathcal{R}^\$)^\dagger = \mathcal{R}^\#$. Now, the action of the $\mathcal{T}$ symmetry (Eq.~\ref{Tsymmetry}) implies $\mathcal{R}^\#=\mathcal{R}$, $H^\#_B=H_B$ and $H^\#_F = H_F = H_F^\$$. Hamiltonians of this type are {\it symplectically similar} to a diagonal matrix i.e. diagonalizable by real symplectic matrices $W\in Sp(M,\mathds{R}): W^\$=W^{-1}$~\cite{de2016diagonalizability}. Consequently, the SVD flattened matrix in this class obeys  
\begin{equation}
 Q=\mathcal{U}\mathcal{V}^T  \in Sp(M,\mathds{R}).
 \label{simplQ}
\end{equation}
Like the orthogonal case, symplectic $Q$ matrices also have distinct homotopy maps compared to both complex and real matrices. Indeed all three, unitary, orthogonal and symplectic are topologically distinct manifolds.

Among commuting symmetries, we need to consider only one antiunitary symmetry because the collection of all such symmetries can be factored into the unitary set and a set that are products of a unitary symmetry and $\mathcal{T}$~\cite{dyson1962threefold}. So the above completes a discussion of antiunitary symmetries that commute with $\mathcal{R}$.

Turning to ``anticommuting'' unitary symmetries, they act on $\mathcal{R}$ like
\begin{equation}
\mathcal{S}: \mathcal{R} \to -\mathcal{S}_F\mathcal{R}\mathcal{S}_B^\dagger.
\label{Ssymmetry}
\end{equation}
The square of $\mathcal{S}$ is a commuting symmetry and working in a basis where this symmetry is diagonal, we see that $\mathcal{S}$ is also block diagonal in this basis for $[\mathcal{S}^2,\mathcal{S}]=0$. So we can restrict ourselves to working within a block where $\mathcal{S}^2 =e^{i\phi} I$ and then further restrict ourselves to the case where $\phi=0$ by just multiplying $\mathcal{S}$ by $e^{-i\phi/2}$. Hence, for the purposes of classification, we can focus on $\mathcal{S}^2 = I$. The eigenvalues of $\mathcal{S}$ are then $\pm 1$ and in its eigenbasis we can write $\mathcal{S}_F =\sigma^F_3$, $\mathcal{S}_B = \sigma^B_3$ where the $\sigma_3$'s could have a different number of 1s than -1s on its diagonal.

Now, an $\mathcal{R}$ matrix satisfying an $\mathcal{S}$ symmetry of the above form is block-off diagonal. The bosonic Hamiltonian $H_B$, on the other hand, is block diagonal for it commutes with $\mathcal{S}_B$. The fermionic Hamiltonian $H_F$ is also block off-diagonal assuming $\sigma_3^F\sigma_2\sigma_3^F = -\sigma_2$ is canonical up to a swap of position and momentum variables. This suggests we can re-express the bosonic problem in terms of a new rigidity matrix $\mathcal{R}'$ with the same $H_B$ but where $\mathcal{S}$ is a commuting symmetry. Indeed, such a map follows from
\begin{equation}
  \mathcal{R} = \begin{pmatrix} 0 & B\\ C & 0\end{pmatrix} = 
\begin{pmatrix} 0 & I\\ I & 0\end{pmatrix}\begin{pmatrix} C & 0\\ 0 & B\end{pmatrix} \equiv
\sigma^F_1\mathcal{R}'.
\end{equation}
Only the auxiliary fermion Hamiltonian changes under this map for $H_F = \sigma_1H_F'\sigma_1$ where $H_F'$ is the fermion Hamiltonian for a system with rigidity matrix $\mathcal{R}'$. So we can always replace an anticommuting symmetry $\mathcal{S}$ with a commuting version and it doesn't represent a different class of symmetry transformations on rigidity matrices. 

Finally, turning to anticommuting antiunitary symmetries (i.e. chiral symmetries) $\mathcal{C}$ with $\mathcal{C}^2=\pm1$, we see we can always write $\mathcal{C} = \mathcal{S}\mathcal{T}$ where $\mathcal{S}$ is an anticommuting unitary symmetry of the type discussed above and $\mathcal{T}$ is a commuting antiunitary symmetry also of the type discussed above. Then, following the above discussion for $\mathcal{S}$, we can map to a rigidity matrix that commutes with $\mathcal{S}$ and for it $\mathcal{C}$ becomes another commuting antiunitary symmetry. So again, we have not found another class of symmetries and we conclude that rigidity matrices follow a three-fold way rather than the full ten-fold way classification.

The three cases discussed above are summarized in the following table.
\begin{center}
\begin{tabular}{ |c|c|c|c|c| } 
 \hline
 Case & Symmetry & $\mathcal{T}^2=$ & Action on $\mathcal{R}$ & Space of $Q$ \\
 \hline
 Complex & $\mathcal{T}=0$ & 0 & $\mathcal{R}=\mathcal{R}$ & $U(M)$ \\ 
 Real & $\mathcal{T}=\mathcal{K}$ & +1 & $\mathcal{R}=\mathcal{R}^*$ & $O(M)$ \\ 
 Symplectic & $\mathcal{T}=\iota\sigma_2\mathcal{K}$ & -1 & $\mathcal{R}=\mathcal{R}^\#$ & $Sp(M,\mathds{R})$ \\ 
 \hline
\end{tabular}
\label{table1}
\end{center}  

\noindent
The topological invariants associated with the rigidity matrices belonging to the three classes are given by the homotopy maps of the classifying spaces of $Q$ which we note down below. From the Bott periodicity theorem on classical groups~\cite{bott1959stable} we obtain the following table till the seventh homotopy group
\begin{center}
\begin{tabular}{ |c|c|c|c|c|c|c|c|c|c| } 
 \hline
 Case       & $\mathcal{T}^2=$ & $\pi_0$ & $\pi_1$ & $\pi_2$ & $\pi_3$ & $\pi_4$ & $\pi_5$ & $\pi_6$ & $\pi_7$ \\
 \hline
 Complex    & 0  & 0 & $\mathds{Z}$ & 0 & $\mathds{Z}$ & 0 & $\mathds{Z}$ & 0 & $\mathds{Z}$ \\ 
 Real       & +1 & $\mathds{Z}_2$ & $\mathds{Z}_2$ & 0 & $\mathds{Z}$ & 0 & 0 & 0 & $\mathds{Z}$ \\ 
 Symplectic & -1 & 0 & 0 & 0 & $\mathds{Z}$ & $\mathds{Z}_2$ & $\mathds{Z}_2$ & 0 & $\mathds{Z}$ \\ 
 \hline
\end{tabular}
\end{center}
after which the elements repeat. In the latter sections, by exemplifying certain frustrated magnets, we will illustrate the implications of some of these invariants in regard to understanding the robust nature of frustration from topological concepts.   

\subsection{Three-fold way for $\nu\neq0$ systems}

In distinction to the $\nu=0$ systems discussed above, the classifying space of the rigidity matrices for a $|\nu|\neq0$ system is a member of the Stiefel manifold (both the $\nu$ and $-\nu$ cases belonging to the same manifold) and naturally falls beyond the ten-fold classification of Hamiltonian matrices. The homotopy groups of different Stiefel manifolds are quite exotic~\cite{saito1955homotopy, james1958intrinsic, matsunaga1959homotopy, gilmore1967complex, mori1972homotopy, james1976topology} supporting invariants other than $\mathbb{Z}$ or $\mathbb{Z}_2$ rendering an exclusive topology to the $\nu\neq0$ frustrated systems.  

For any $\nu\neq0$ system, there are three different structures of the Stiefel manifold defined over either a real, complex or quaternionic (symplectic) space. For example, the complex Stiefel manifold $V_M(\mathds{C}^N)$ is the set of all $M$-tuples ($x_1, . . . , x_M$) of orthonormal vectors in $\mathds{C}^N$. While the definition translates to other vector spaces as well (the real analog is explained in Fig.~\ref{fig:manifold}), we start with the complex case first.

For a generic $\nu\neq 0$ system, $\mathcal{R}$ is a random complex rectangular matrix. There exists a couple of different definitions of SVD of a rectangular matrix. We adopt the one where the SVD of a $M\times N$ complex matrix implies $\mathcal{R}=\mathcal{U}\Lambda_R \mathcal{V}^{\dagger}$ with $\mathcal{U}\in U(M)$, $\mathcal{V}\in U(N)$ and $\Lambda_R$ is a diagonal $M\times N$ matrix. For example, if $M<N$ i.e. $\nu\equiv N-M>0$ ($N$ d.o.fs and $M$ constraints), flattening the elements of $\Lambda_R$ yields
\begin{equation}
\Lambda_R \rightarrow \tilde{\Lambda}_R =   \left(
	      \begin{array}{rrrr|rrr}
		  1 & 0 & \cdots & 0 & 0 & \cdots & 0 \\
		  0 & 1 & \cdots & 0 & 0 & \cdots & 0 \\
		  \vdots & \vdots & \vdots & \vdots & \vdots & \vdots & \vdots \\
		  \undermat{M}{0 & 0 & \cdots & 1} & \undermat{N-M}{0 & \cdots & 0} \\
	      \end{array}
	      \right).  \nonumber \\
	      \vspace{5mm}
\end{equation}
In this case, $Q=\mathcal{U}\tilde{\Lambda}_R \mathcal{V}^{\dagger}$ changes when the transformation is a non-trivial element of $[U(M)\times U(N)]/[U(M)\times U(N-M)]\cong U(N)/U(N-M)$, namely the complex Stiefel manifold $V_M(\mathds{C}^N)$ whose homotopy groups dictate the topology of frustration described by $\mathcal{R}$. 

For $\nu=1$ i.e. $M=N-1$, the space $V_{N-1}(\mathds{C}^N)$ is diffeomorphic to the classical group $SU(N)$ whose homotopy maps $\pi_d$ are isomorphic to that of $U(N)$ for $d\ge2$ (which is the same as the $\nu=0$ case) while $\pi_1[SU(N)]=0$. For other $\nu\ge2$, we take note of the homotopy exact sequence $\pi_{d-1}[V_{N-\nu}(\mathds{C}^N)]\cong \pi_{d-1}[S^{2\nu+1}]$ which leads to the following table
 \begin{center}
 \begin{tabular}{ |c|c|c|c|c|c|c|c|c|c|c| } 
 \hline
  $\nu$ & $\pi_1$ & $\pi_2$ & $\pi_3$ & $\pi_4$ & $\pi_5$ & $\pi_6$ & $\pi_7$ & $\pi_8$ & $\pi_9$ & $\pi_{10}$ \\
 \hline
    1 & 0 & 0 & $\mathds{Z}$ & 0 & $\mathds{Z}$ & 0 & $\mathds{Z}$ & 0 & $\mathds{Z}$ & 0  \\
 \hline
    2 & 0 & 0 & 0 & 0 & $\mathds{Z}$ & $\mathds{Z}_{2}$ & $\mathds{Z}_2$ & $\mathds{Z}_{24}$ & $\mathds{Z}_2$ & $\mathds{Z}_2$ \\ 
 \hline
    3 & 0 & 0 & 0 & 0 & 0 & 0 & $\mathds{Z}$ & $\mathds{Z}_2$ & $\mathds{Z}_2$ & $\mathds{Z}_{24}$ \\
 \hline
    4 & 0 & 0 & 0 & 0 & 0 & 0 & 0 & 0 & $\mathds{Z}$ & $\mathds{Z}_2$ \\
 \hline
    5 & 0 & 0 & 0 & 0 & 0 & 0 & 0 & 0 & 0 & 0 \\
 \hline 
 \end{tabular}
 \end{center} 
Note there is no particular periodicity like the $\nu=0$ case. So we continue the table until for a given $\nu$, all homotopy maps are trivial in dimension $d=0$ to $d=10$. One pronounced implication of the table is that complex rigidity matrices corresponding to $\nu\neq 0$ systems has nontrivial topological features only in a close neighborhood of the $\nu=0$ point (this is also true for the real and the symplectic cases as we will see soon). This is one of the important results of this paper which we interpret in the following way. So far topology has been explored mostly in $\nu=0$ systems, however, $\nu\neq 0$ systems as well can display discerning topological signatures and indeed there exist ample examples of frustrated magnets which advocate for the statement.

In presence of the $\mathcal{T}$-symmetry which makes $\mathcal{R}$ real, the classifying space of $Q=\mathcal{U}\tilde{\Lambda}_R \mathcal{V}^T$ is the real Stiefel manifold $V_M(\mathds{R}^N)$. Like the complex case, for $\nu=1$, $V_M(\mathds{R}^N)$ is diffeomorphic to the classical group $SO(N)$. Since $SO(N)$ is the identity component of $O(N)$, all their homotopy groups after $\pi_0$ match with $\pi_0[SO(N)]=0$ and $\pi_1[SO(2)]=\mathds{Z}$~\cite{dodson1997user}. The homotopy exact sequence $\pi_{d-1}[V_{N-\nu}(\mathds{R}^N)]\cong \pi_{d-1}[S^{\nu}]$ implies $\pi_d[V_{N-\nu}(\mathds{R}^N)]=0$ if $d<\nu$, using which we obtain the following table 
 \begin{center}
 \begin{tabular}{ |c|c|c|c|c|c|c|c|c|c|c| } 
 \hline
  $\nu$ & $\pi_1$ & $\pi_2$ & $\pi_3$ & $\pi_4$ & $\pi_5$ & $\pi_6$ & $\pi_7$ & $\pi_8$ & $\pi_9$ & $\pi_{10}$ \\
 \hline
    1 & $\mathds{Z}_2$ & 0 & $\mathds{Z}$ & 0 & 0 & 0 & $\mathds{Z}$ & $\mathds{Z}_2$ & $\mathds{Z}_2$ & 0  \\
 \hline
    2 & 0 & $\mathds{Z}$ & $\mathds{Z}$ & $\mathds{Z}_2$ & $\mathds{Z}_2$ & $\mathds{Z}_{12}$ & $\mathds{Z}_2$ & $\mathds{Z}_2$ & $\mathds{Z}_3$ & $\mathds{Z}_{15}$ \\ 
 \hline
    3 & 0 & 0 & $\mathds{Z}$ & $\mathds{Z}_2$ & $\mathds{Z}_2$ & $\mathds{Z}_{12}$ & $\mathds{Z}_2$ & $\mathds{Z}_2$ & $\mathds{Z}_3$ & $\mathds{Z}_{15}$ \\
 \hline
    4 & 0 & 0 & 0 & $\mathds{Z}$ & $\mathds{Z}_2$ & $\mathds{Z}_{2}$ & $\mathds{Z}_2\times\mathds{Z}_{12}$ & $\mathds{Z}_2^2$ & $\mathds{Z}_2^2$ & $\mathds{Z}_{24}\times\mathds{Z}_{3}$ \\
 \hline
    5 & 0 & 0 & 0 & 0 & $\mathds{Z}$ & $\mathds{Z}_{2}$ & $\mathds{Z}_2$ & $\mathds{Z}_{24}$ & $\mathds{Z}_2$ & $\mathds{Z}_2$ \\
 \hline
    6 & 0 & 0 & 0 & 0 & 0 & $\mathds{Z}$ & $\mathds{Z}_2$ & $\mathds{Z}_2$ & $\mathds{Z}_{24}$ & 0 \\
 \hline
    7 & 0 & 0 & 0 & 0 & 0 & 0 & $\mathds{Z}$ & $\mathds{Z}_2$ & $\mathds{Z}_2$ & $\mathds{Z}_{24}$ \\
 \hline
    8 & 0 & 0 & 0 & 0 & 0 & 0 & 0 & $\mathds{Z}$ & $\mathds{Z}_2$ & $\mathds{Z}_2$ \\
 \hline
    9 & 0 & 0 & 0 & 0 & 0 & 0 & 0 & 0 & $\mathds{Z}$ & $\mathds{Z}_2$ \\
 \hline 
   10 & 0 & 0 & 0 & 0 & 0 & 0 & 0 & 0 & 0 & $\mathds{Z}$ \\
 \hline
   11 & 0 & 0 & 0 & 0 & 0 & 0 & 0 & 0 & 0 & 0 \\
 \hline
 \end{tabular}
 \end{center} 

For the $\mathcal{T}$-symmetry with $\mathcal{T}^2=-1$, a quaternionic (symplectic) Stiefel manifold $V_M(\mathds{H}^N)$ describes the classifying space of $Q$. When $\nu=0$, we recover the $\nu=0$ case as $V_N(\mathds{H}^N)\cong Sp(N)$. For other $\nu$, the homotopy exact sequence $\pi_{d-1}[V_{N-\nu}(\mathds{H}^N)]\cong \pi_{d-1}[S^{4\nu+3}]$ leads to the following table
 \begin{center}
 \begin{tabular}{ |c|c|c|c|c|c|c|c|c|c|c| } 
 \hline
  $\nu$ & $\pi_1$ & $\pi_2$ & $\pi_3$ & $\pi_4$ & $\pi_5$ & $\pi_6$ & $\pi_7$ & $\pi_8$ & $\pi_9$ & $\pi_{10}$ \\
 \hline
    1 & 0 & 0 & 0 & 0 & 0 & 0 & $\mathds{Z}$ & $\mathds{Z}_{2}$ & $\mathds{Z}_2$ & $\mathds{Z}_{24}$ \\ 
 \hline
    2 & 0 & 0 & 0 & 0 & 0 & 0 & 0 & 0 & 0 & 0 \\
 \hline 
 \end{tabular}
 \end{center}
In this case, the homotopy groups of $Q$ and the topology of $\mathcal{R}$ become trivial starting from $\nu=2$ and onward in dimensions up to $d=10$. 

When $\mathcal{R}$ is a square matrix, flattening of the singular values leads to $\Lambda_R\rightarrow\tilde{\Lambda}_R=\mathbb{I}$ only if all the singular values are nonzero, in other words, $\mathcal{R}$ has a full rank. If the dimension of $\mathcal{R}$ is $N$ while its rank is $M<N$, there will be $N-M$ zero diagonal entries in $\Lambda_R$. In that case, the matrix $\tilde{\Lambda}_R\mathcal{V}^T$ will have only $M$ linearly independent vectors. We can then add $N-M$ random vectors to $\tilde{\Lambda}_R\mathcal{V}^T$ and make all of them orthonormal by following the Gram-Schmidt process to transform $Q$ into the appropriate class of matrices based on the symmetries of $\mathcal{R}$. This way, the topology of a $\nu=0$  system with $M$ zero modes per unit cell (i.e. $M$ zero diagonal entries in $\Lambda_R$) is effectively described by that of a $\nu\neq 0$ system with $|\nu|=M$. Later we will illustrate this situation by exemplifying certain types of kagome antiferromagnets which host a flat band of zero modes by virtue of a high degree of frustration. 

Finally, we emphasize an important mathematical point that underlies the classification. The rigidity matrices need to be defined continuously across the parameter space of interest. For example, in the case of periodic ground states, the rigidity matrix should be a continuous function of ${\bf k}$. Only when this continuity is present does it make sense to talk about the topology of the rigidity matrices. In general, starting from a bosonic Hamiltonian $H_B=\mathcal{R}^\dagger\mathcal{R}$ and taking its square root would not lead to a rigidity matrix $\mathcal{R}$ that is continuous in its parameter space. We achieve continuity in our examples by writing the Hamiltonian as a set of constraints of the form $H_{\rm meta}$ (in Eq.~\ref{Hmetamat}) and only then deriving the rigidity matrix by expanding about a ground state. Regardless of the approach, the classification applies whenever the rigidity matrices vary continuously across their parameter space.  

Having realized the classification table of non-Hermitian matrices under the action of time-reversal symmetry, we now elucidate the topological invariants that characterize the topology of the zero modes arising from the low dimensional homotopy groups up to $\pi_2$. This includes nodal points, lines or surfaces of zero modes realized in certain frustrated magnet spin-wave band structures that are describable within the theory of rigidity matrices presented above. Perhaps, a simpler starting point is to consider the ensembles of random unitary and orthogonal matrices as the spaces generically feature topologically protected zero modes of the above-mentioned forms. The nodal points or the Weyl points are protected by an integer winding number while the nodal lines similar to the Fermi surface of a metal are characterized by a $\mathbb{Z}_2$ invariant at each point in the space of matrices. In the orthogonal ensemble, we further find a new vortex-like invariant manifesting as $\mathbb{Z}_2$ strings.

\section{Examples of topological invariants}\label{secfive}

In this section, we provide the details of constructing the topological invariants which characterize the zero modes in both $\nu=0$ and $\nu\neq 0$ systems. The forms of the zero modes of our concern typically include points, lines, or two-dimensional surfaces for which the appropriate homotopy map to look at is $\pi_0$, $\pi_1$, and $\pi_2$ respectively. We consider the ensembles of random unitary, orthogonal, or symplectic matrices as the representatives of those systems and construct the appropriate topological invariant for each of these classes.      

\subsection{$\nu=0$ systems}

For complex rigidity matrices, the corresponding SVD flattened matrix $Q$ is unitary. According to the table, a nontrivial topology associated to $U(N)$ can only arise in odd dimensions, starting from $\pi_1$. As the determinant of a unitary matrix is unimodular, the argument of Det($Q$) can have nontrivial winding in the space of unitary matrices. Such winding guarantees the existence of zero modes in the form of Weyl points. The homotopy group $\pi_1$ encodes the topology of loops in this space giving rise to an invariant in the form of winding number~\cite{kane2014topological} 
\begin{equation}
 w = \frac{1}{2\pi} \oint_{\Gamma} d~{\rm arg}[{\rm Det}(Q)] \in\mathbb{Z},
 \label{topinv1}
\end{equation}
where $\Gamma$ is a close contour surrounding a Weyl point. So, $\pi_1$ of $U(N)$ implies the possibilities of realizing topological zero modes characterized by a $\mathbb{Z}$ invariant in $\nu=0$ systems described by a complex rigidity matrix. A simple example of this kind is the balls-springs systems discussed in Ref.~[\onlinecite{kane2014topological}] where the boundary modes are protected by a nonzero value of the winding number $w$. 

The SVD flattened matrix $Q$ corresponding to a real rigidity matrix is orthogonal, hence, has determinant $\pm1$. The table for $O(N)$ suggests that each point in the space of orthogonal matrices has a nontrivial topology given by $\pi_0$ which we can characterize by a topological invariant 
\begin{equation}
 \eta = {\rm Det}(Q) \in\mathbb{Z}_2.
 \label{topinv2}
\end{equation}
In the space of $O(N)$ there are regions of determinant with $1$ and $-1$ separated by lines of zero modes across which $\eta$ changes sign. These are similar to the Fermi surface of metals in two dimensions. One encounters this situation in certain classes of KHAFs where the line nodes are protected by a $\mathbb{Z}_2$ topology~\cite{roychowdhury2017spin}. Later we will furnish other examples of frustrated magnets where frustration is protected by the topology of real rigidity matrices.  

\begin{figure}
\centering    
\includegraphics[width=0.8\columnwidth]{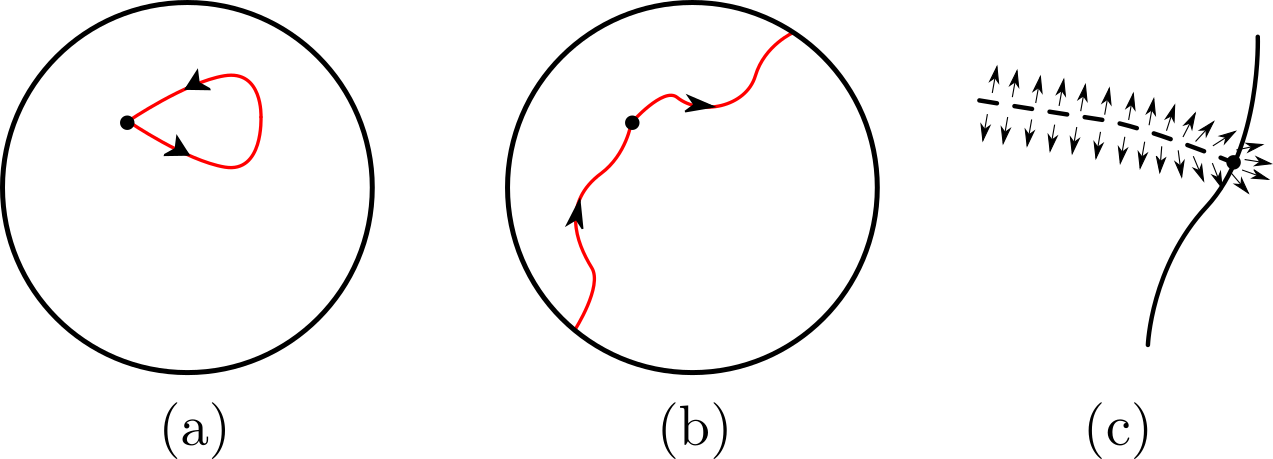}
\caption{(a) A contractible loop on the $SO(3)$ sphere. (b) A noncontractible loop on the $SO(3)$ sphere. (c) The Dirac string emanating from a $\mathds{Z}_2$ vortex which lies on the line node. The vector ${\bf n}$ (described in the text) flips across the string.}
\label{fig:z2string}
\end{figure}

Besides $\pi_0$, the homotopy group $\pi_1$ of $O(N)$ is also $\mathbb{Z}_2$. To demonstrate $\pi_1$ which encodes the topology of loops in these spaces, we consider the well known example of $SO(3)$, the rotation group in three dimensions. The meaning of $\pi_1[SO(3)]=\mathbb{Z}_2$ is that all closed loops in the space of $SO(3)$ fall into two homotopy classes -- those that are contractible [Fig.~\ref{fig:z2string} (a)] and those that are not [Fig.~\ref{fig:z2string} (b)]. Composing two paths from the second class yields a path from the first class. The origin of the noncontractibility in the second class can be understood as the following. Each element of $SO(3)$ can be specified geometrically by an axis ${\bf n}$ (a three-dimensional unit vector) and a rotation angle $\theta$ about ${\bf n}$ with the redundancy that rotations of $\theta=\pi$ around ${\bf n}$ and $-{\bf n}$ are identical. So, we can visualize the situation as a solid sphere of radius $\pi$ with a one-to-one correspondence between the points in the sphere and the elements of $SO(3)$ -- the position vector of any point $P$ is specified by the direction set by ${\bf n}$ and the magnitude set by $\theta$. Having said that, the sphere has the antipodal points on the surface identified, thus, for $\theta$ larger than $\pi$, the point $P$ appears on the other side of the sphere. Accordingly, the vector ${\bf n}$ flips its direction as it hits any antipodal point. 

\begin{figure}
\centering    
\includegraphics[width=1.0\columnwidth]{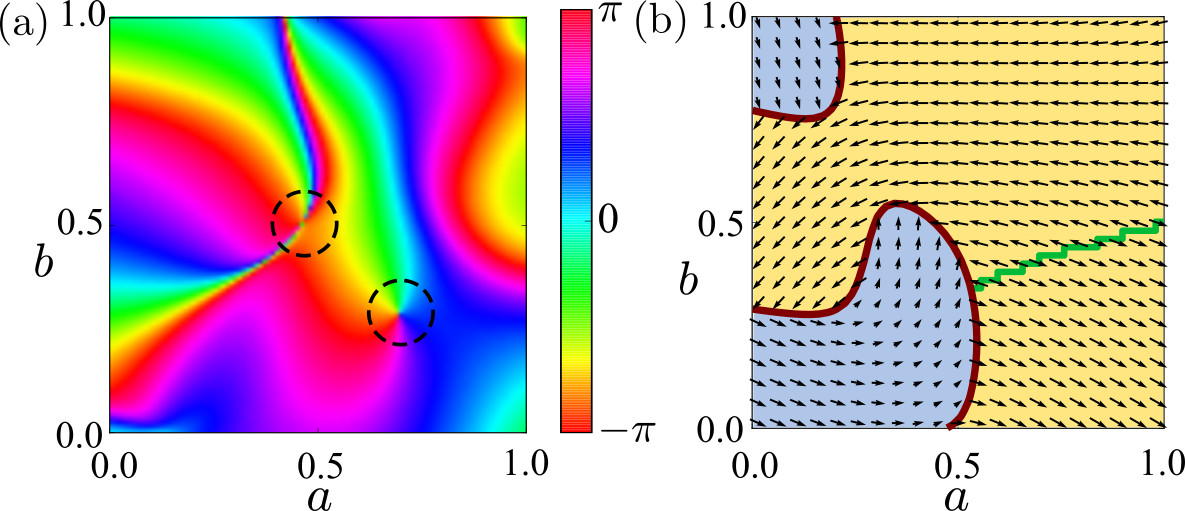}
\caption{(a) When $\mathcal{R}$ is a random complex matrix, its SVD flattening yields $Q\in U(N)$ (Eq.~\ref{Qunit}), and so, a plot of arg[Det($\mathcal{R}$)] reveals Weyl points with winding number $w\in\mathbb{Z}$ in the parameter space specified by $a$ and $b$ (defined in Eq.~\ref{randR}) which are marked by dashed black circles. Here we consider $Q\in U(3)$. (b) When $\mathcal{R}$ is a random real matrix, the SVD flattening yields $Q\in O(N)$ (Eq.~\ref{Qorth}). Line nodes arise when the determinant of $Q$ changes. The yellow and the blue region in the parameter space represent $\eta$ (in Eq.~\ref{topinv2}) to be $+1$ and $-1$ respectively separated by line nodes (shown in dark red). The green line represents a Dirac string as a signature of the associated $\mathbb{Z}_2$ topology that emanates from a point on the line node and across the string the vector ${\bf n}$ (its $x$ and $y$ components are shown by black arrows) explained in the text flips.}
\label{fig:randmat}
\end{figure}

Now consider a closed path in the space of real $\mathcal{R}$ matrices or equivalently the orthogonal matrices obtained from the SVD flattening of $\mathcal{R}$. If we trace ${\bf n}$ on the  sphere as one traverses along the closed path, and observe it to flip sign, we have found a Dirac string. The matrices $\mathcal{R}$ changed smoothly but our description of $\mathcal{Q}$ changed abruptly. If this closed path is noncontractible, this must happen an odd number of times. The topological invariant that characterizes a close path $\Gamma$ in the space of $SO(3)$ matrices is then given by  
\begin{equation}
 \zeta = (-1)^{N} \in \mathds{Z}_2,
 \label{topinv3}
\end{equation}
where $N$ is the number of times $\Gamma$ crosses the Dirac string(s) [i.e. number of times ${\bf n}$ changed sign as in Fig.~\ref{fig:z2string} (c)]. The arguments carry {\it mutatis mutandis} to other $SO(N)$ matrices with $N>3$ for which ${\bf n}$ becomes a $N(N-1)/2$-dimensional unit vector (since a $SO(N)$ matrix can be parametrized by a set of $N(N-1)/2$ independent parameters).

The above results reveal the topology of $\nu=0$ systems which we further illustrate by performing numerical simulations over ensembles of square rigidity matrices. In doing so we consider four random matrices $\mathcal{R}_{(1,2,3,4)}$ as the four corners of a square grid specified by two variables $a,b\in[0,1]$. Any point on this square grid is given by 
\begin{equation}
\mathcal{R}=(1-a)(1-b)\mathcal{R}_1+(1-a)b\mathcal{R}_2+ab\mathcal{R}_3+a(1-b)\mathcal{R}_4. 
\label{randR}
\end{equation}
For the complex case, the plot of arg[Det($\mathcal{R}$)] over the square evidences the existence of Weyl points with integer winding $w$ arising from $\pi_1$ (Eq.~\ref{topinv1}) as shown in Fig.~\ref{fig:randmat} (a). A similar plot for the real case [Fig.~\ref{fig:randmat} (b)] features line nodes owing to $\pi_0$ (Eq.~\ref{topinv2}) and Dirac strings associated to $\pi_1$ (Eq.~\ref{topinv3}). Fig.~\ref{fig:randmat} (b) further suggests that the Dirac strings emanates from a $\mathds{Z}_2$ vortex that lies on the line nodes [see also Fig.~\ref{fig:z2string}.(c)]. In the symplectic case, the determinant of the SVD flattened matrix $Q$ (see Eq.~\ref{simplQ}) associated with $\mathcal{R}$ in Eq.~\ref{fig:randmat} is 1~\cite{rim2017elementary}, hence, the topology is trivial for both $\pi_0$ and $\pi_1$.

\subsection{$\nu\neq0$ systems}

So far we have discussed the topology of the $\nu=0$ systems but an extension to $\nu\neq0$ ones is straightforward. For further elucidation on the topology in the latter class of models let us take the $|\nu|=1$ systems as an example. The classifying spaces of the SVD flattened matrices $Q$ for $|\nu|=1$ are the subgroups of the corresponding classical groups as mentioned before. The real case is of specific interest to us for there are frustrated magnets belonging to this class in which the zero modes are protected by the topology coming from $\pi_1$ and also that the lower homotopy groups for the complex and the symplectic cases are trivial for $|\nu|=1$ systems. 

\begin{figure}
\centering    
\includegraphics[width=1.0\columnwidth]{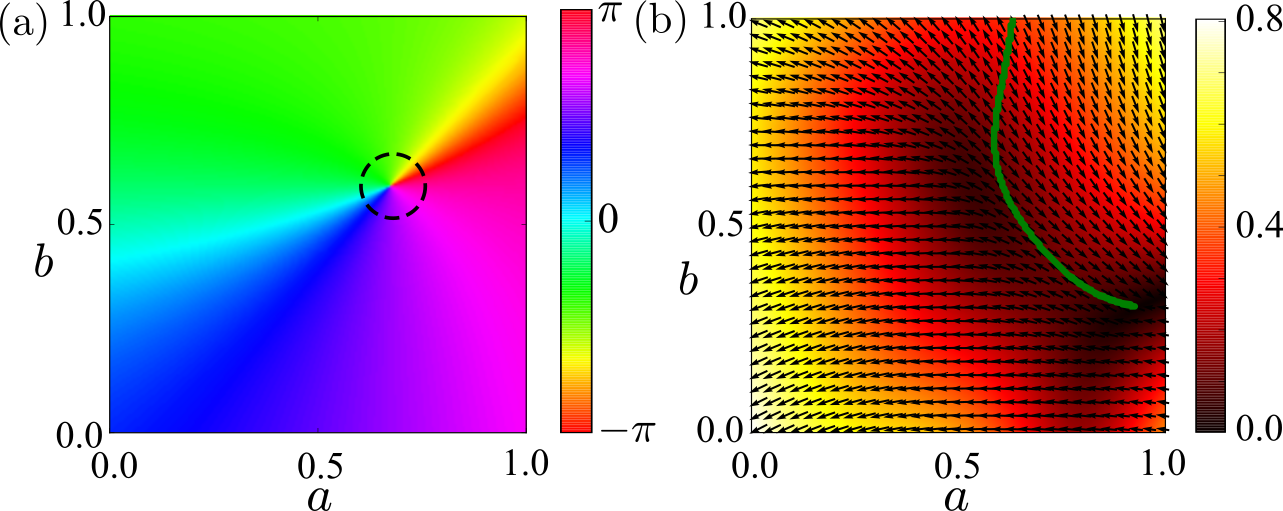}
\caption{(a) When $\mathcal{R}$ is a random real matrix with $\nu=1$ and has $1\times2$ structure, its topology is decided by the parameter $\theta$ in Eq.~\ref{Qtheta1} a plot of which reveals Weyl points with winding number $w\in\mathbb{Z}$ in the parameter space specified by $a$ and $b$ (defined in Eq.~\ref{randR}); one such point is located and marked by dashed black circles. (b) When $\mathcal{R}$ is a random real matrix with $\nu=1$ but has $2\times3$ structure, its SVD flattening yields a $SO(3)$ matrix, so like the $\nu=0$ case shown in Fig.~\ref{fig:randmat} we plot the corresponding vector ${\bf{n}}$ (its $x$ and $y$ components shown in black arrows) to reveal the presence of $\mathbb{Z}_2$ vortices. These vortices emerge from or terminate to wherever the lowest singular value of $\mathcal{R}$ touches 0 which we show in the density plot.}
\label{fig:randnu1}
\end{figure}

To illustrate the topology in $|\nu|=1$ systems, we first consider the class of real $1\times2$ matrices. The SVD flattening of $\mathcal{R}$ belonging to this class leads to  
\begin{equation}
 Q=\begin{pmatrix} \cos\theta & \sin\theta \\ -\sin\theta & \cos\theta \end{pmatrix} \in SO(2),
 \label{Qtheta1}
\end{equation}
whose first nontrivial homotopy group is $\pi_1$ and that $\pi_1[SO(2)]=\mathbb{Z}$ which features Weyl points arising from the winding of $\theta$ in the parameter space. We can plot the result in Fig.~\ref{fig:randnu1} (a) for a random ensemble of real $1\times2$ matrices as done previously for the $\nu=0$ case. Later we will illustrate this situation with a classic model of frustrated system in which the full rigidity matrix decouples into small $1\times2$ blocks each featuring such $\mathbb{Z}$ topology that protects the zero modes in that system.  

Next we consider the ensemble of real $2\times3$ matrices. The SVD of $\mathcal{R}$ belonging to this ensemble implies that $\mathcal{U}\in O(2)$, $\mathcal{V}\in O(3)$, while flattening of the singular values yields $\tilde{\Lambda}_R=\begin{pmatrix} 1 & 0 & 0 \\ 0 & 1 & 0 \end{pmatrix}$. A triad of three vectors ${\bf q}_{(1,2,3)}$ can be formed first by orthonormalizing the two rows of $Q=\mathcal{U}\tilde{\Lambda}_R\mathcal{V}^T$ to form two vectors ${\bf q}_1$ and ${\bf q}_2$, and then construct ${\bf q}_3=({\bf q}_1\times {\bf q}_2)/|{\bf q}_1\times {\bf q}_2|$. This way we can map $Q$ to an $SO(3)$ matrix whose rows are given by ${\bf q}_{(1,2,3)}$. The rest of the analysis of topology then follows from the arguments presented in the previous subsection regarding $SO(3)$ which yields the plot presented in Fig.~\ref{fig:randnu1} (b). So, $\zeta$ (in Eq.~\ref{topinv3}) is defined for this system and allows for $Z_2$ vortices, but $\eta = {\rm Det}[\mathcal{Q}\to SO(3)]=1$ and so there are no line nodes. An important feature to note here is that the $Z_2$ vortices, in this case, emanate from or terminate to wherever the lowest singular value of $\mathcal{R}$ touches 0 [see the density plot in Fig.~\ref{fig:randnu1} (b)] reducing its rank further by 1 at that particular point.

In passing, let us also briefly mention the scenario for $|\nu|=2$ systems. The ensemble of real matrices that have dimensions $3\times1$ (or $1\times3$) forms the simplest possible example. In this case, the SVD flattened matrix $Q$ represents a unit vector ($\hat{\bf d}$) in three dimensions. The first nontrivial topology in this case comes from $\pi_2$ which is the homotopy maps of closed surfaces and characterized by the integer-valued topological invariant given by the Chern-Pontryagin index~\cite{kitaev2006anyons} 
\begin{equation}
 \mathcal{P} = \frac{1}{4\pi} \iint \hat{\bf d}\cdot(\partial_{x^\alpha}\hat{\bf d}\times\partial_{x^\beta}\hat{\bf d}) dx^\alpha dx^\beta \in \mathds{Z}.
 \label{chern}
\end{equation}

As in the $\nu=0$ case, here we show the existence of zero modes associated with the $\nu\neq 0$ topology using random matrices. Again, generated complex, real and symplectic varieties, we generate two dimensional images as shown in Fig.~\ref{fig:randnu1}. These demonstrate the above topological invariants and show without requiring an understanding of homotopy groups that the $\nu\neq 0$ cases also have zero modes demanded by topological invariants. 

\section{Examples of frustration by topological invariants}\label{secsix}

In order to investigate how signatures of frustration acquire robustness owing to topology, we consider two classic examples of frustrated magnetic systems --\\ 1) the $J_1-J_2$ Heisenberg model on a square lattice and \\ 2) kagome Heisenberg antiferromagnets with a flat band. \\ The robustness is verified by introducing perturbations that break the spin rotation symmetry at various levels. While the former has found much significance in the study of High-T$_{\rm C}$ superconductivity in certain cuprates and iron-based compounds~\cite{xu2008ising, ma2008arsenic, si2008strong}, the latter offers a fertile ground of realizing new exotic states of matter such as spin liquid~\cite{lacroix2011introduction, diep2013frustrated, savary2016quantum}.

\begin{figure}
\centering    
\includegraphics[width=1.0\columnwidth]{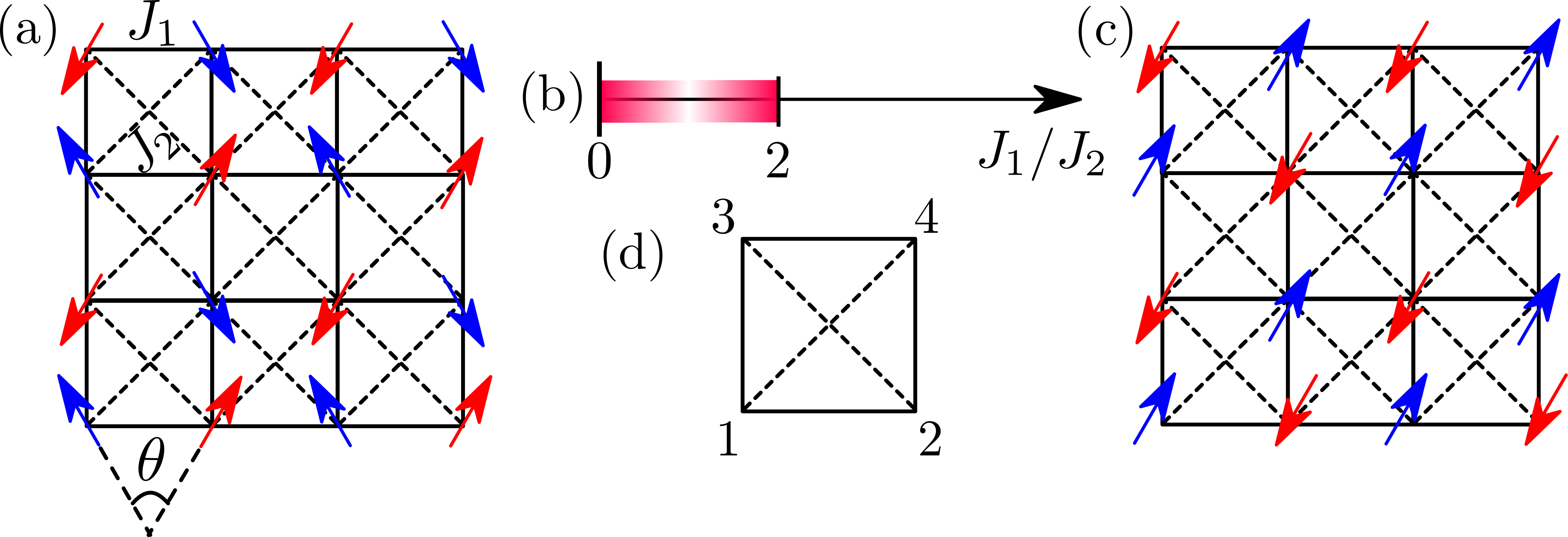}
\caption{(a) A typical ground state spin configuration of the $J_1-J_2$ model in the frustrated regime which occurs for $J_1/J_2<2$ [shaded in the phase diagram in (b)]. The ground state decouples into two sublattices (one with red spins and the other with blue spins) each having a N\'{e}el order, however, the relative angle $\theta$ between the spins in each of them can be arbitrary. For $J_1/J_2>2$, a N\'{e}el state preponderates over the entire lattice as shown in (c). The constraints in the model live on the small square plaquettes one of which is shown in (d) with the spins enumerated on which the LT transformation described in the text applies.}
\label{fig:j1j2pd}
\end{figure}

\subsection{The $J_1-J_2$ model on a square lattice}

The $J_1-J_2$ Heisenberg model on a square lattice is specified by the Hamiltonian~\cite{chandra1990ising, misguich2013two}
\begin{equation}
 H = J_1\sum_{\langle i,j\rangle} {\bf S}_i\cdot{\bf S}_j + J_2\sum_{\langle\langle i,j\rangle\rangle} {\bf S}_i\cdot{\bf S}_j,
 \label{j1j2ham}
\end{equation}
where $\langle i,j\rangle$ and $\langle\langle i,j\rangle\rangle$ denote the spin pairs of nearest and next-nearest neighbors (nnn) with antiferromagnetic interaction $J_1$ and $J_2$ respectively ($J_{1,2}>0$). An extensive amount of work can be found in the literature depicting the phase diagram (see Fig.~\ref{fig:j1j2pd}) of the model at low temperatures (see Ref.~[\onlinecite{misguich2013two}] for a review and the references therein). Following the Luttinger-Tisja (LT) theorem~\cite{luttinger1946theory} (see Also Ref.~[\onlinecite{litvin1974luttinger}] and the references therein), the classical energy can be minimized by a helical spin texture ${\bf S}_i = \hat{{\bf e}}_1 \cos({\bf q}\cdot{\bf r}_i) + \hat{{\bf e}}_2 \sin({\bf q}\cdot{\bf r}_i)$, where the wave vector ${\bf q}$ minimizes the Fourier transform $J({\bf q})$ of the coupling in Eq.~\ref{j1j2ham}. For $J_2<J_1/2$, the minimum of $J({\bf q})$ is achieved at ${\bf q}=(\pi,\pi)$ featuring a N\'{e}el ordering of the spins. At the critical point $J_2=J_1/2$ the model is highly frustrated as $J({\bf q})$ has lines of minima around the edges of the square BZ. For $J_2>J_1/2$, the minima localize to ${\bf q}=(\pi,0)$ and ${\bf q}=(0,\pi)$. This is also a frustrated state (however, the degree of frustration is less than the critical point) resulting from a decoupling of the two sublattices each having a N\'{e}el order, however, the relative angle ($\theta$) between the spins in each of them can be arbitrary. The result is a degenerate manifold of ground states parametrized by the continuous angle $\theta$ (see Fig.~\ref{fig:j1j2pd}). At the critical point, this entire manifold becomes degenerate with the N\'{e}el state. 

\begin{figure*}
\centering    
\includegraphics[width=1.9\columnwidth]{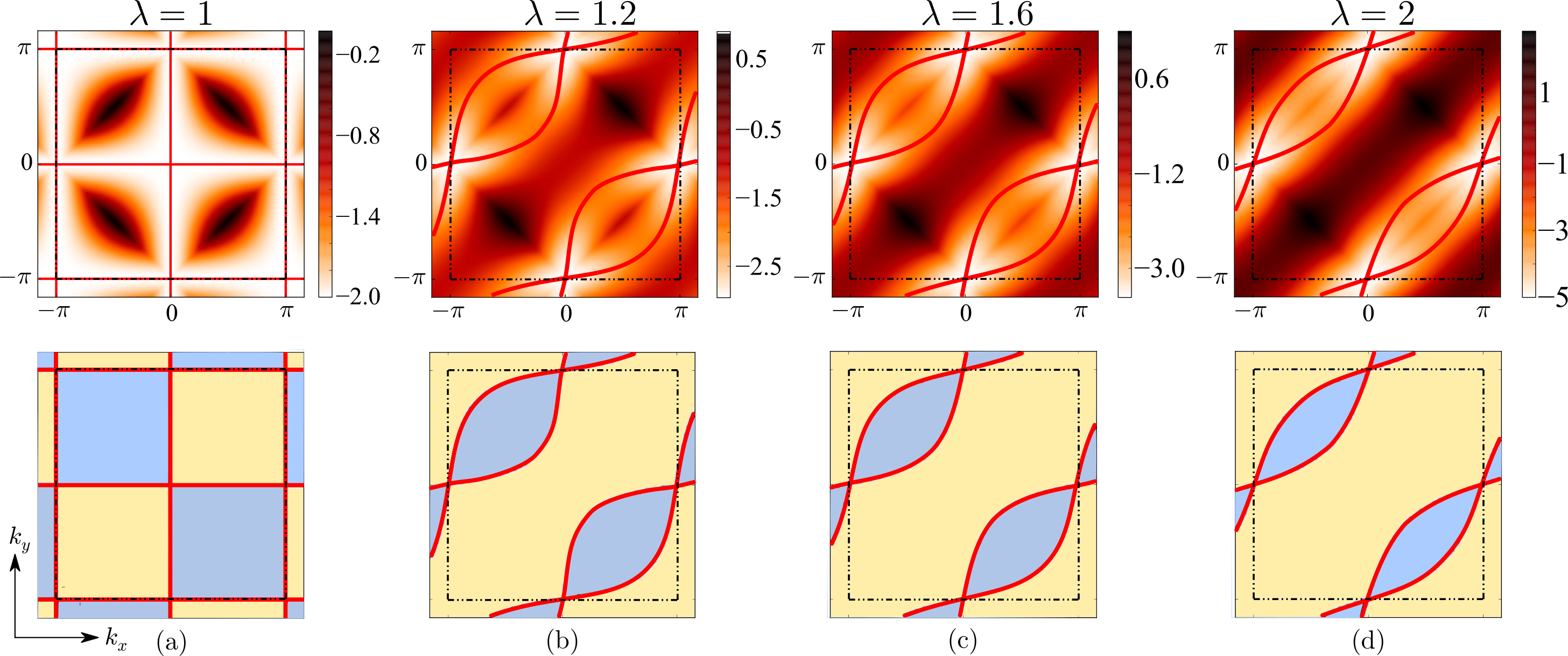}
\caption{Top row shows the plots of the LT spectrum over BZ for various values of the diagonal anisotropy $\lambda$ defined in Eq.~\ref{j1j2ham3} where $\lambda=1$ implies absence of such anisotropies and reduces Eq.~\ref{j1j2ham3} to Eq.~\ref{j1j2ham}. The bottom row displays the plots of the topological invariant $\eta={\rm sign}[{\rm Det}(\tilde{\mathcal{R}})]$ (defined in Eq.~\ref{etacrit}) for $\lambda=1$ (no anisotropies) in (a) and $\eta={\rm sign}[r_1r_2]$ (defined in Eq.~\ref{r1r2}) for other values of $\lambda$ in (b)-(d). The yellow and the blue region have $\eta=+1$ and $\eta=-1$ respectively. The plots evince how Weyl line nodes tend to pairwise merge to Dirac line nodes~\cite{roychowdhury2017spin} for large anisotropy.}
\label{fig:j1j2crit}
\end{figure*}

In order to identify the constraints associated with this model it is useful to reexpress the Hamiltonian in Eq.~\ref{j1j2ham} as
\begin{equation}
  H=\left\{
  \begin{array}{@{}ll@{}}
    \sum_{\Square} H_{\Square}^+, & \text{if}\ g>0 \\
    \sum_{\Square} H_{\Square}^-, & \text{otherwise},
  \end{array}\right.
  \label{j1j2cond}
\end{equation} 
where
\begin{equation}
\begin{split} 
 H_{\Square}^\pm = \frac{J_1}{4} \bigg[ ({\bf S}_{1}&+{\bf S}_{2}+{\bf S}_{3}+{\bf S}_{4})^2 \\ 
 &+ 2|g| \big\{({\bf S}_{1}\pm{\bf S}_{3})^2+({\bf S}_{2}\pm{\bf S}_{4})^2\big\} \bigg],
\end{split}
\label{j1j2ham2}
\end{equation}
with $g=(J_2/J_1-1/2)$, and $\Square$ indexing a square plaquette comprising four spins ${\bf S}_{1,2,3,4}$ in a counterclockwise direction (see Fig.~\ref{fig:j1j2pd}). Without loss of generality, we can assume the direction of the collinear order along $+x$ axis in the spin space so that we can linearize the fluctuations around ${\bf S}_{i}=(1,0,0)$ and to linear order, write ${\bf S}_{i}=(1,q^{i},p_{i})$. Now we are all set to construct the rigidity matrix of the problem, however, its shape depends on the sign of the dimensionless parameter $g$. Let us address the three different cases for $g<0$, $g=0$, and $g>0$ separately to make clear distinctions. They correspond to the N\'{e}el ordered state (for $g\leq0$) and the frustrated region (for $g>0$) respectively.

\subsubsection{The N\'{e}el state for $g<0$}

For $g<0$, the energy is minimized by the N\'{e}el state (Fig.~\ref{fig:j1j2pd}). Following the condition in Eq.~\ref{j1j2cond}, we need to consider the Hamiltonian $H_{\Square}^-$ in Eq.~\ref{j1j2ham2} which has total nine constraints per unit cell in the ground state. They are $L_1^\alpha\equiv\sum_{i\in \Square}S_{i}^\alpha=0$, $L_2^\alpha\equiv S_{1}^\alpha-S_{3}^\alpha=0$, and $L_3^\alpha\equiv S_{2}^\alpha-S_{4}^\alpha=0$ with $\alpha\in\{x,y,z\}$ which we expand around a N\'{e}el ordered state. For the particular spin configuration we choose for the N\'{e}el state, all constraints corresponding to $\alpha=x$ contribute only to vanishing rows of $\mathcal{R}$, making it effectively a $6\times2$ matrix. 

To obtain a translation invariant Bravais lattice corresponding to the N\'{e}el pattern for which the LT theorem applies, we perform the following transformation on the spins in one of the sublattices, namely
\begin{equation}
 S_{2,4}^{x,y} \rightarrow - \tilde{S}_{2,4}^{x,y} ~~;~~ S_{2,4}^{z} \rightarrow \tilde{S}_{2,4}^{z}.
\end{equation}
We call this transformed basis the {\it LT basis}, in which the rigidity matrix $\mathcal{R}$ takes the desired $6\times2$ form. In other words, in each unit cell we have the LT basis: $\tau_1= [q_1,p_1]^T$ and $\tau_2=[L_1^y,L_2^y,L_3^y,L_1^z,L_2^z,L_3^z]^T$, such that $\tau_2=\mathcal{R}\cdot\tau_1$ with the Hamiltonian (Eq.~\ref{j1j2ham}) in the transformed basis written as
\begin{equation}
 H_{LT} = -J_1\sum_{\langle i,j\rangle} (S^x_i\tilde{S}^x_j+S^y_i\tilde{S}^y_j-S^z_i\tilde{S}^z_j) + J_2\sum_{\langle\langle i,j\rangle\rangle} {\bf S}_i\cdot\tilde{\bf S}_j.
 \label{j1j2ham1}
\end{equation}

In the momentum space, $\mathcal{R}$ acquires a block diagonal form as
\begin{equation}
 \mathcal{R}({\bf k}) =
 \begin{pmatrix}
  1-z_x-z_y+z_xz_y & 0 \\
  1-z_xz_y & 0 \\
  z_x-z_y & 0 \\
  0 & 1+z_x+z_y+z_xz_y \\
  0 & 1-z_xz_y \\
  0 & z_x-z_y 
 \end{pmatrix}.
 \label{R_neel2}
\end{equation}
Two antiunitary symmetries $\mathcal{T}_1\equiv\mathcal{K}\mathcal{C}_1$ and $\mathcal{T}_2\equiv\mathcal{K}\mathcal{C}_2$ ($\mathcal{K}$ denotes the complex conjugation) with
\begin{equation}
 \mathcal{C}_1({\bf k}) =
 \begin{pmatrix}
  1 & 0 & 0 & 0 & 0 & 0 \\
  0 & -1 & 0 & 0 & 0 & 0 \\
  0 & 0 & -1 & 0 & 0 & 0 \\
  0 & 0 & 0 & 1 & 0 & 0 \\
  0 & 0 & 0 & 0 & -1 & 0 \\
  0 & 0 & 0 & 0 & 0 & -1
 \end{pmatrix},
 \label{R_neel2}
\end{equation}
and $\mathcal{C}_2({\bf k}) = z_xz_y\mathbb{I}_{2\times2}$ satisfy $\mathcal{T}^\dagger_1({\bf k})\mathcal{R}({\bf k})\mathcal{T}_2({\bf k})=\mathcal{R}({\bf k})$ which serves as a symmetry of $\mathcal{R}$ in the momentum space (the unitary parts correspond to a $\tilde{\mathcal{C}}_2$-rotation symmetry i.e. rotation of 180$^\circ$ around the center of the square plaquette in the N\'{e}el state). So in the basis of these antiunitary symmetries, $\mathcal{R}$ has real elements as 
\begin{equation}
 \mathcal{R}({\bf k}) =
 \begin{pmatrix}
  -4f_1\sin\frac{k_x}{2}\sin\frac{k_y}{2} & 0 \\
  2|\sin\frac{k_x+k_y}{2}| & 0 \\
  -2f_2\sin\frac{k_x-k_y}{2} & 0 \\
  0 & 4f_1\cos\frac{k_x}{2}\cos\frac{k_y}{2} \\
  0 & 2|\sin\frac{k_x+k_y}{2}| \\
  0 & -2f_2\sin\frac{k_x-k_y}{2}
 \end{pmatrix},
 \label{R_neel3}
\end{equation}
where $f_1={\rm sign}(\cos\frac{k_x+k_y}{2})$ and $f_2={\rm sign}(\sin\frac{k_x+k_y}{2})$. Denoting $\mathcal{R}({\bf k}) \equiv \begin{pmatrix} \mathcal{R}_1({\bf k}) & 0 \\ 0 & \mathcal{R}_2({\bf k}) \end{pmatrix}$, both the blocks represent systems with $\nu=2$ for which the lowest homotopy group with a nontrivial topology is $\pi_2$. Both of them have vanishing elements at certain points in the BZ, however, not simultaneously, namely, $\mathcal{R}_1({\bf k})$ vanishes at ${\bf k}=(0,0)$ while $\mathcal{R}_2({\bf k})$ vanishes at ${\bf k}=(\pm\pi,\pm\pi)$. For this reason, it is not possible to define the Chern-Pontryagin index for any of them that could reveal the topology of $\pi_2$ in this case. This leads us to conclude that the unfrustrated N\'{e}el state for $g<0$ is not topology protected. 

\subsubsection{The critical point at $g=0$}

At the critical point ($g=0$), we have only three constraints in the problem, namely $L^\alpha\equiv\sum_{i\in \Square}S_{i}^\alpha=0$ with $\alpha\in\{x,y,z\}$ in the ground state. Again we expand these constraints around a N\'{e}el ordered state which qualifies as one of the many ground states at the this point. Translated to the momentum space, $\mathcal{R}$ assumes the following form written in the LT basis $\tau_1= [q_1,p_1]^T$ and $\tau_2= [L^x,L^y,L^z]^T$,
\begin{equation}
 \mathcal{R}({\bf k}) =
 \begin{pmatrix}
  0 & 0 \\
  1-z_x-z_y+z_xz_y & 0 \\
  0 & 1+z_x+z_y+z_xz_y
 \end{pmatrix},
 \label{R_neel}
\end{equation}
where $z_{x,y}=e^{ik_{x,y}}$. By using the symmetries of the spin pattern together with the crystal symmetries as in the previous subsection, one can find appropriate antiunitary symmetries $\mathcal{T}_1=\mathcal{K}$ and $\mathcal{T}_2=z_xz_y\mathcal{T}_1$ such that $\mathcal{T}^\dagger_1({\bf k})\mathcal{R}({\bf k})\mathcal{T}_2({\bf k})=\mathcal{R}({\bf k})$, and $\mathcal{R}$ expressed in the basis of these antiunitary symmetries, has real elements as 
\begin{equation}
 \mathcal{R}({\bf k}) =
 \begin{pmatrix}
  0 & 0 \\
  -4f\sin\frac{k_x}{2}\sin\frac{k_y}{2} & 0 \\
  0 & 4f\cos\frac{k_x}{2}\cos\frac{k_y}{2}
 \end{pmatrix},
 \label{R_neel}
\end{equation}
where $f={\rm sign}(\cos\frac{k_x+k_y}{2})$, and the vanishing row corresponds to the constraint $L^x$. We find that $J({\bf k})$ corresponding to $H_{LT}$ has lines of minima at $k_{x,y}=0$ and $\pm\pi$ (Fig.~\ref{fig:j1j2crit} (a) top panel), and the quantity ${\rm sign}[{\rm Det}(\tilde{\mathcal{R}})]$, where $\tilde{\mathcal{R}}$ is the diagonal matrix derived from $\mathcal{R}$ by eliminating the vanishing row, changes sign across the lines (Fig.~\ref{fig:j1j2crit} (a) bottom panel). Thus, the zero modes in this case are protected by the $\mathbb{Z}_2$ invariant 
\begin{equation}
 \eta={\rm sign}[{\rm Det}(\tilde{\mathcal{R}})]
 \label{etacrit}
\end{equation}
(compare with Eq.~\ref{topinv2}) which arises from $\pi_0$ of the real case in the table of $\nu=0$ systems. This is reminiscent of the line nodes observed in Ref.~[\onlinecite{roychowdhury2017spin}] protected by a $\mathbb{Z}_2$ topology.

\subsubsection{The frustrated state for $g>0$}

\begin{figure}
\centering    
\includegraphics[width=1.0\columnwidth]{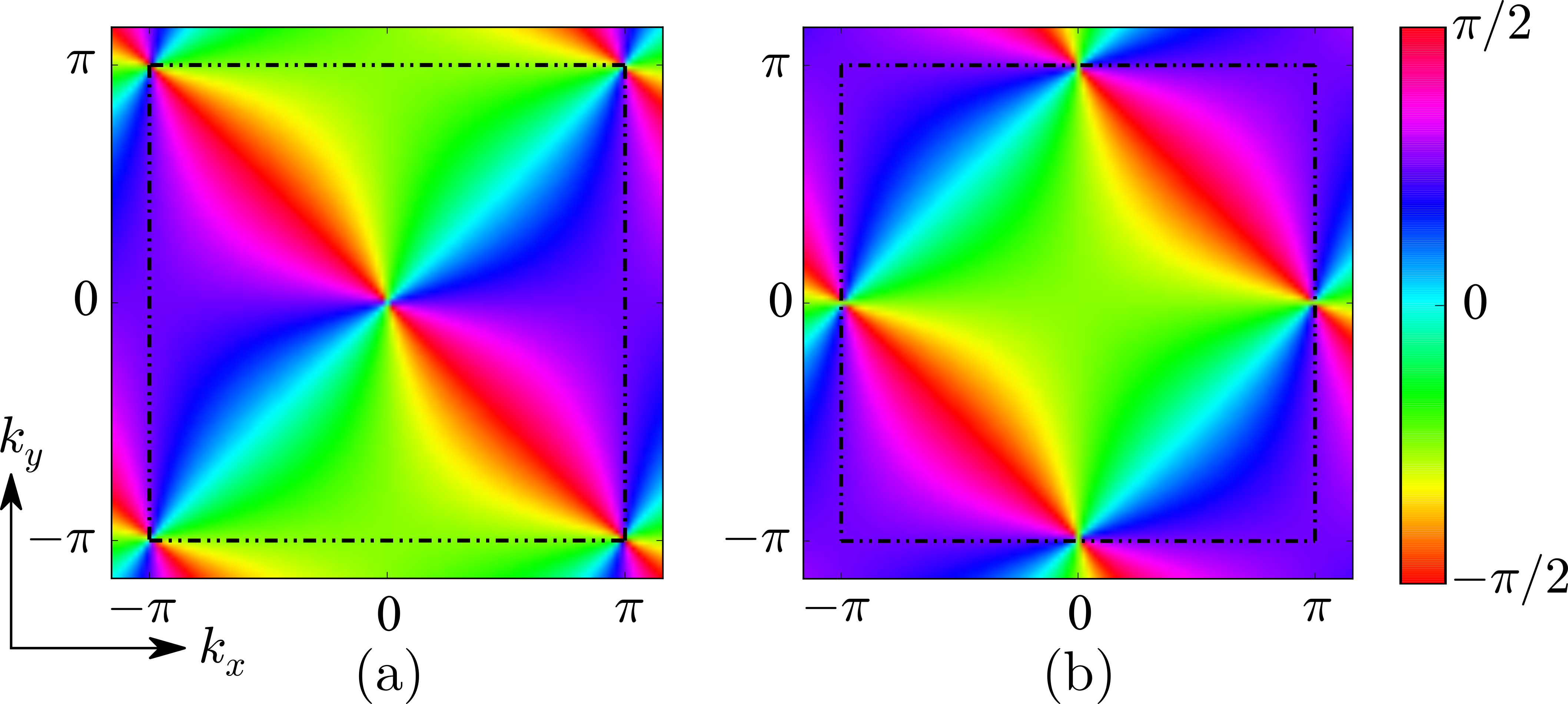}
\caption{A plot of (a) $\theta_1$ and (b) $\theta_2$ defined in Eq.~\ref{Qtheta} over the BZ (with the zone boundary marked in dashed black lines) for $g=0.1$ (with $J_1=1$) and a particular spin configuration with $\theta=1$. The Weyl points located at ${\bf k}=(0,0)$ and ${\bf k}=(\pm\pi,\pm\pi)$ for $\theta_1$ and at ${\bf k}=(\pm\pi,0)$ and ${\bf k}=(0,\pm\pi)$ for $\theta_2$ signify the feature of topology protected frustration for $g>0 $.}
\label{fig:j1j2frust}
\end{figure}

For $g>0$, frustration is attributed to the two sublattices individually conceiving N\'{e}el order and the energetics of the model being insensitive to the relative angle ($\theta$) between them (Fig.~\ref{fig:j1j2pd}). So the accidental degeneracy of the ground states is specified by the continuous parameter $\theta$. The LT transformation for such a spin pattern with a given value of $\theta$ would be
\begin{equation}
\begin{split}
  &S_{1}^{x,y,z} \rightarrow \tilde{S}_{1}^{x,y,z} \\ 
  &S_{2}^{x} \rightarrow \cos\theta\tilde{S}_{2}^{x}+\sin\theta\tilde{S}_{2}^{y} \\ 
  &S_{2}^{y} \rightarrow -\sin\theta\tilde{S}_{2}^{x}+\cos\theta\tilde{S}_{2}^{y} ~~;~~ S_2^z \rightarrow \tilde{S}_2^z \\  
  &S_{3}^{x,y} \rightarrow -\tilde{S}_{3}^{x,y} ~~;~~ S_3^z \rightarrow \tilde{S}_3^z \\ 
  &S_{4}^{x} \rightarrow -\cos\theta\tilde{S}_{4}^{x}-\sin\theta\tilde{S}_{4}^{y} \\ 
  &S_{4}^{y} \rightarrow \sin\theta\tilde{S}_{4}^{x}-\cos\theta\tilde{S}_{4}^{y} ~~;~~ S_4^z \rightarrow \tilde{S}_4^z,
\end{split}
\end{equation}
using which it is straightforward to construct $H_{LT}$ (the complicated expression is not provided here). For further analysis, we resort to this LT transformed basis only. Considering the Hamiltonian  $H_{\Square}^+$ in Eq.~\ref{j1j2cond} and Eq.~\ref{j1j2ham2} for $g>0$, we find total nine constraints per unit cell in the ground state which are $L_1^\alpha\equiv\sum_{i\in \Square}S_{i}^\alpha=0$, $L_2^\alpha\equiv S_{1}^\alpha+S_{3}^\alpha=0$, and $L_3^\alpha\equiv S_{2}^\alpha+S_{4}^\alpha=0$ with $\alpha\in\{x,y,z\}$. Evidently, not all of them are linearly independent. After a careful elimination of all the dependent constraints we find $\mathcal{R}$ reduced to a $4\times2$ matrix which, in the momentum space, assumes the form 
\begin{equation}
 \mathcal{R}({\bf k}) =
 \begin{pmatrix}
  1-z_xz_y & 0 \\
  z_x- z_y & 0 \\
  0 & 1+z_xz_y \\
  0 & z_x+z_y 
 \end{pmatrix}.
 \label{R_frust2}
\end{equation}
The antiunitary symmetries $\mathcal{T}_{1,2}\equiv\mathcal{K}\mathcal{C}_{1,2}$ which constitute the symmetry of $\mathcal{R}$ as $\mathcal{T}^\dagger_1({\bf k})\mathcal{R}({\bf k})\mathcal{T}_2({\bf k})=\mathcal{R}({\bf k})$ have $\mathcal{C}_1({\bf k})=\mathbb{I}_{4\times4}$ and $\mathcal{C}_2({\bf k})=z_xz_y \begin{pmatrix} -1 & 0 \\ 0 & 1 \end{pmatrix}$. The corresponding real form of $\mathcal{R}$ is
\begin{equation}
 \mathcal{R}({\bf k}) =
 \begin{pmatrix}
  2|\sin\frac{k_x+k_y}{2}| & 0 \\
  -f_2\sin\frac{k_x-k_y}{2} & 0 \\
  0 & 2|\cos\frac{k_x+k_y}{2}| \\
  0 & f_1\cos\frac{k_x-k_y}{2} 
 \end{pmatrix},
 \label{R_frust3}
\end{equation}
where $f_1={\rm sign}(\cos\frac{k_x+k_y}{2})$ and $f_2={\rm sign}(\sin\frac{k_x+k_y}{2})$. Let us denote $\mathcal{R}({\bf k}) \equiv \begin{pmatrix} \mathcal{R}_1({\bf k}) & 0 \\ 0 & \mathcal{R}_2({\bf k}) \end{pmatrix}$. The SVD flattening of $\mathcal{R}_{1,2}({\bf k})$ leads to two $SO(2)$  matrices 
\begin{equation}
 Q_{1,2}=\begin{pmatrix} \cos\theta_{1,2} & \sin\theta_{1,2} \\ -\sin\theta_{1,2} & \cos\theta_{1,2} \end{pmatrix}
 \label{Qtheta}
\end{equation}
for which the lowest homotopy group with a nontrivial topology is $\pi_1$, and that $\pi_1[SO(2)]=\mathbb{Z}$. This is evident from the plots of $\theta_{1,2}$ over the BZ shown in Fig.~\ref{fig:j1j2frust} in which we observe Weyl points at ${\bf k}=(0,0)$ and ${\bf k}=(\pm\pi,\pm\pi)$ for $\theta_1$ and at ${\bf k}=(\pm\pi,0)$ and ${\bf k}=(0,\pm\pi)$ for $\theta_2$. Thus, the frustrated state for $g>0$ in the $J_1-J_2$ model is actually protected by a $\mathbb{Z}$ topology.

\subsubsection{Inclusion of diagonal anisotropies}

The topologically protected zero modes are immune to certain classes of perturbations made to the Hamiltonian in Eq.~\ref{j1j2ham}. One of them is the diagonal anisotropies in which one of the diagonal interactions ($J_2$) is stronger than the other. The model has the following Hamiltonian
\begin{equation}
 H = J_1\sum_{\langle i,j\rangle} {\bf S}_i\cdot{\bf S}_j + \lambda J_2\sum_{\langle\langle i,j\rangle\rangle_{d_1}} {\bf S}_i\cdot{\bf S}_j + \frac{J_2}{\lambda}\sum_{\langle\langle i,j\rangle\rangle_{d_2}} {\bf S}_i\cdot{\bf S}_j,
 \label{j1j2ham3}
\end{equation}
where $\langle\langle i,j\rangle\rangle_{d_1}$ and $\langle\langle i,j\rangle\rangle_{d_2}$ denote the two diagonal interactions weighted by a dimensionless factor of $\lambda$ and $\lambda^{-1}$ respectively (note the symmetry $\lambda\leftrightarrow1/\lambda$). We can cast this Hamiltonian in an analogous form to Eq.~\ref{j1j2cond} where 
\begin{equation}
\begin{split}  
 H_{\Square}^\pm = \frac{J_1}{4} \bigg[ ({\bf S}_{1}&+\lambda{\bf S}_{2}+{\bf S}_{3}+\lambda{\bf S}_{4})^2/\lambda \\ 
 &+ 2|g| \big\{({\bf S}_{1}\pm{\bf S}_{3})^2/\lambda+\lambda({\bf S}_{2}\pm{\bf S}_{4})^2\big\} \bigg],
\end{split}
\label{j1j2ham4}
\end{equation}
obtaining constraints same as before except $L_1^\alpha$ modifies to $L_1^\alpha\equiv S_{1}^\alpha+\lambda S_{2}^\alpha+S_{3}^\alpha+\lambda S_{4}^\alpha=0$. The effects of this perturbation at different parts of the phase diagram are the following
\begin{itemize}
 \item For $g<0$, we need to consider all the constraints given by $L^\alpha_{1,2,3}$ acting on the N\'{e}el state, consequently Eq.~\ref{R_neel3} modifies to
 \begin{equation}
  \mathcal{R}({\bf k}) =
  \begin{pmatrix}
   r_1 & 0 \\
   2|\sin\frac{k_x+k_y}{2}| & 0 \\
   -2f_2\sin\frac{k_x-k_y}{2} & 0 \\
   0 & r_2 \\
   0 & 2|\sin\frac{k_x+k_y}{2}| \\
   0 & -2f_2\sin\frac{k_x-k_y}{2}
  \end{pmatrix}.
  \label{R_neel4}
 \end{equation}
 For $\lambda$ away from 1, the Chern-Pontryagin index ($\mathcal{P}$) is well defined for $\mathcal{R}_1$ and $\mathcal{R}_2$. However, we only get $\mathcal{P}=0$ indicating a trivial topology associated with the unfrustrated N\'{e}el state.
 \item At the critical point i.e. $g=0$, only $L_1^\alpha$ with $\alpha\in\{y,z\}$ contribute. The rigidity matrix takes the form 
 \begin{equation}
  \mathcal{R} = 
  \begin{pmatrix}
   r_1 & 0 \\
   0 & r_2 
  \end{pmatrix}, 
 \end{equation}
 where
 \begin{equation}
  \begin{split}
   r_1 = 2f\bigg(\cos\frac{k_x+k_y}{2}-\lambda\cos\frac{k_x-k_y}{2}\bigg), \\
   r_2 = 2f\bigg(\cos\frac{k_x+k_y}{2}+\lambda\cos\frac{k_x-k_y}{2}\bigg),
  \end{split}
  \label{r1r2}
 \end{equation}
 with $f={\rm sign}[\cos\frac{k_x+k_y}{2}]$. As we tune $\lambda$ away from 1 (the two ranges $0<\lambda<1$ and $\lambda\geq1$ are mappable by $\lambda\rightarrow1/\lambda$), we observe changes in the locations of the lines of zero modes which are characterized by the $\mathbb{Z}_2$ invariant $\eta={\rm sign}[r_1r_2]$ [Fig.~\ref{fig:j1j2crit} (b)-(d)]. For a high value of $\lambda$, pairwise merging of the lines leads to doubly degenerate line nodes along the $k_x-k_y=\pm\pi$ lines in the BZ. These are Dirac type of line nodes, distinct from the singly degenerate Weyl type of line nodes. Both the kinds were earlier reported in the Ref.~[\onlinecite{roychowdhury2017spin}]. The Dirac line nodes are also protected by a $\mathbb{Z}_2$ topology, however, with a new topological invariant which is $\eta={\rm sign}[r_{1}]$ or $\eta={\rm sign}[r_{2}]$. 
 \item For $g>0$, the scenario does not change from the unpurterbed case of $\lambda=1$ since $\lambda$ does not enter in the expression of $\mathcal{R}$. We can conclude that the frustrated state is robust against this kind of diagonal perturbations. For large $\lambda$, we effectively get a triangular lattice which in fact favors the frustrated state keeping its topology invariant.  
\end{itemize} 

\subsubsection{Inclusion of spin rotation symmetry breaking terms}

Perturbations that can induce new constraints to the model, can potentially alter the topology of frustration. Let us investigate the effects of certain spin rotation symmetry breaking terms added to the Hamiltonian in Eq.~\ref{j1j2ham}. These are easy axis anisotropies in the spin space which tend to align the spins in a preferred direction, this way, relieving the frustration and destroying its topology. We consider the following Hamiltonian
\begin{equation}
 H = J_1\sum_{\langle i,j\rangle} {\bf S}_i\cdot{\bf S}_j + \tilde{J_1}\sum_{\langle i,j\rangle} S_i^xS_j^x + J_2\sum_{\langle\langle i,j\rangle\rangle} {\bf S}_i\cdot{\bf S}_j.
 \label{j1j2ham5}
\end{equation}
The second term can be reexpressed as the following
\begin{equation}
\begin{split}
 S_i^xS_j^x = \frac{1}{2}(S_i^x + S_j^x)^2 &+ \frac{1}{2}(S_i^y)^2 + \frac{1}{2}(S_i^z)^2 \\ 
                                           &+ \frac{1}{2}(S_j^y)^2 + \frac{1}{2}(S_j^z)^2 - S^2,
\end{split}
\end{equation}
which introduces new constraints $L^4_h\equiv S_i^x + S_j^x=0$ (on the horizontal nn bond), $L^4_v\equiv S_i^x + S_j^x=0$ (on the vertical nn bond), $L^6\equiv S_i^y=0$, and $L^7\equiv S_i^z=0$ in each unit cell in the ground state. Inclusion of this new set of constraints shifts the system further away from the $\nu=0$ point for which we do not have any nontrivial topology in $\pi_0$, $\pi_1$, or $\pi_2$. In effect, these perturbations destroy the topology of the frustration in the model and gap out the zero modes.

\begin{figure}
\centering    
\includegraphics[width=0.78\columnwidth]{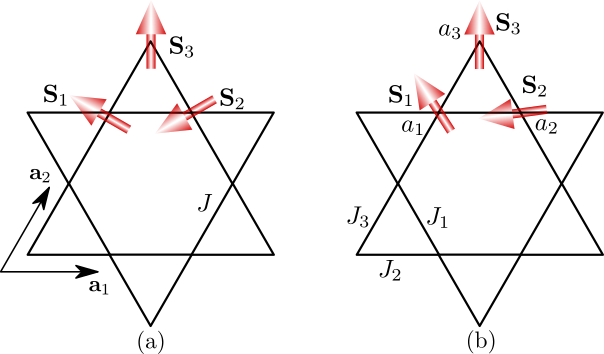}
\caption{(a) Isotropic kagome Heisenberg model (spin exchanges given by $J$) with constraint ${\bf S}_1+{\bf S}_2+{\bf S}_3=0$. The spins are oriented in a 120$^{\circ}$ configuration forming an equilateral triangle. (b) Anisotropic kagome Heisenberg model (spin exchanges given by $J_1$, $J_2$, $J_3$) with constraint $a_1{\bf S}_1+a_2{\bf S}_2+a_3{\bf S}_3=0$. The coefficients $a_{1,2,3}$ are determined by $J_{1,2,3}$ which in turn decide the spin configurations obeying the constraints. The modifications $a_j\rightarrow a_j\mathds{M}_j$ allow for both scalar and spin-orbit type spin exchanges.}
\label{fig:kagome1}
\end{figure}

\subsection{The spin-wave flat band in kagome antiferromagnets}

\begin{figure*}
\centering    
\includegraphics[width=2.0\columnwidth]{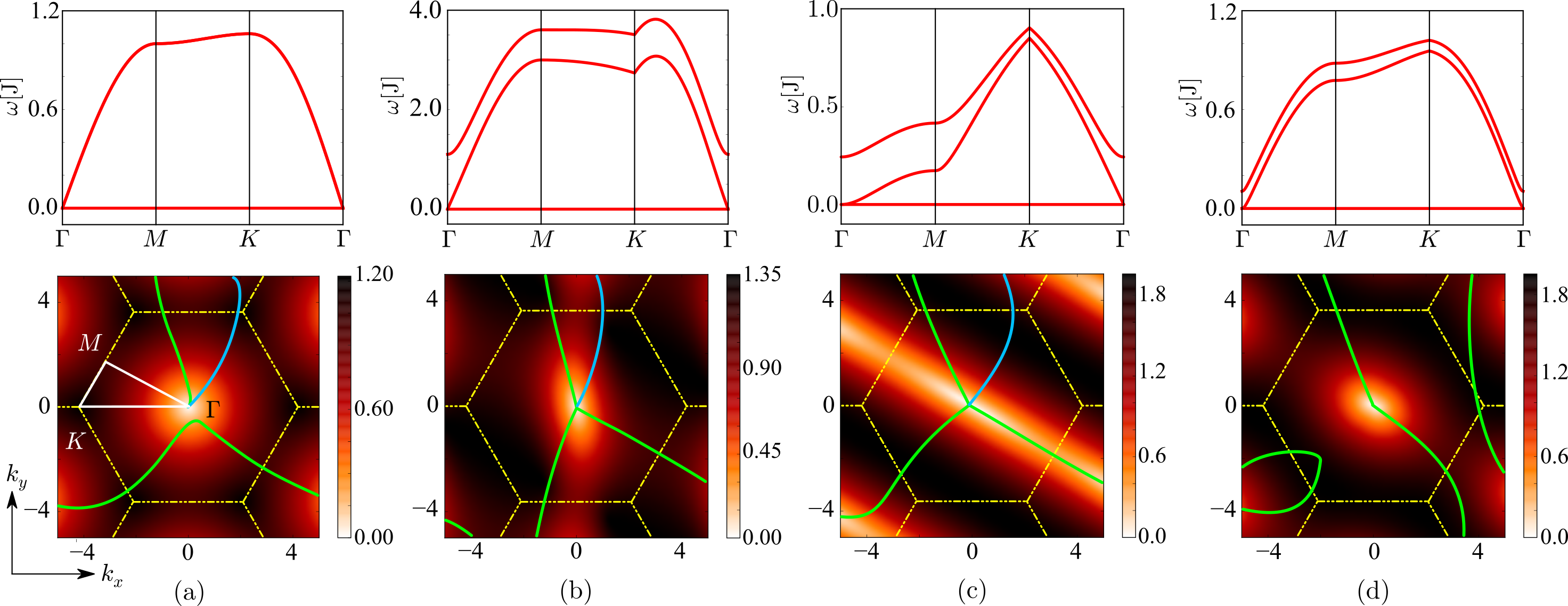}
\caption{Effects of spin-orbit coupling on the rigidity of kagome spin waves. (a) isotropic kagome Heisenberg model (b) anisotropic kagome Heisenberg model (c) spin-orbit coupled kagome antiferromagnets with SO(2) symmetry (Eq.~\ref{Mmat}), and (d) generic spin-orbit coupled kagome antiferromagnets (Eq.~\ref{so3brokenmat} and \ref{so3brokenomega}). The top panel shows the spin-wave band structures in all the four varieties (with parameters mentioned in the text) along a path shown in white lines in the bottom panel of (a). The spin-wave frequencies $\omega$ is measured in units of spin exchange $J$ set to 1. The bottom panel is a plot of the gap between the lowest two singular values of the rigidity matrix $\mathcal{R}$ in the BZ which closes at the $\Gamma$ point from which Dirac strings emanate (green and blue respectively for $\mathcal{R}_1$ and $\mathcal{R}_2$ in Eq.~\ref{rig2} and \ref{rig3}). For the most generic model in (d), the spin-orbit exchanges destroy the block structure of $\mathcal{R}$ and so the nontrivial topology associated with the blocks leaving us with $\zeta=1$.}
\label{fig:kagome2}
\end{figure*}

Kagome antiferromagnets (Fig.~\ref{fig:kagome1}) form a quintessential example of frustrated systems. They can support zero modes in various forms from line or point nodes to flat bands. Let us start with the simplest example of ideal KHAF. The spin Hamiltonian is given in Eq.~\ref{ham2} and the zero energy configurations can be visualized by folding patterns of a triangulated origami sheet~\cite{reimers1993order, shender1993kagome, chandra1993anisotropic, shender1996order}. One prominent candidate for the ground states of the model is the $q=0$ coplanar state [the 120$^{\circ}$ configuration shown in Fig.~\ref{fig:kagome1} (a)] which represents a flat sheet in the origami language. The spin-wave spectrum around this state features a flat band of zero modes [Fig.~\ref{fig:kagome2} (a) top] which turns out to be characterized by a topological invariant $\zeta$ defined in Eq.~\ref{topinv3}.

The unit cell of the $q=0$ spin pattern has three spins i.e. six d.o.fs, and two vector constraints specified by $S_{\triangle\alpha}=0$ with $\alpha\in\{x,y,z\}$ representing a $\nu=0$ system. The rigidity matrix ($\mathcal{R}$) that encodes the fluctuations around the coplanar spin order is a square matrix of dimension six. An explicit construction of $\mathcal{R}$ follows from invoking Eq.~\ref{riggmat} and considering the basis $\tau_1= [q_1,q_2,q_3,p_1,p_2,p_3]^T$ corresponding to the three spins $S_1,S_2,S_3$ in the unit cell and $\tau_2=[\triangle_1^x,\triangle_2^x,\triangle_1^y,\triangle_2^y,\triangle_1^z,\triangle_2^z]^T$ corresponding to the six constraints on the two faces $\triangle_1$ and $\triangle_2$ in the unit cell such that $\tau_2=\mathcal{R}\cdot\tau_1$. In this basis, $\mathcal{R}$ has a block-diagonal form $\mathcal{R}=\begin{pmatrix} \mathcal{R}_1 & 0 \\ 0 & \mathcal{R}_2 \end{pmatrix}$ where $\mathcal{R}_1$ and $\mathcal{R}_2$ are $4\times3$ and $2\times3$ matrices representing two systems of $\nu=-1$ and $\nu=1$ respectively. So the topology of a $\nu=0$ system ($\mathcal{R}$ with $\nu=0$) in this case hinges upon that of the two systems ($\mathcal{R}_{1,2}$) with $\nu=\pm1$ (both having the same topology). In the momentum space, $\mathcal{R}_1$ takes the form
\begin{equation}
 \mathcal{R}_1({\bf k}) = \begin{pmatrix}
                         0 & -\frac{\sqrt{3}}{2} & \frac{\sqrt{3}}{2} \\
                         0 & -\frac{\sqrt{3}}{2} & \frac{\sqrt{3}}{2}e^{ik_1} \\
                         1 & -\frac{1}{2} & -\frac{1}{2} \\
                         e^{ik_2} & -\frac{1}{2} & -\frac{1}{2}e^{ik_1} \\
                        \end{pmatrix},
 \label{rig2}
\end{equation}
and $\mathcal{R}_2$ takes the form
\begin{equation}
 \mathcal{R}_2({\bf k}) = \begin{pmatrix}
                         1 & 1 & 1 \\
                         e^{ik_2} & 1 & e^{ik_1} \\
                        \end{pmatrix},
 \label{rig3}
\end{equation}
where $k_j={\bf k}\cdot{\bf a}_j$ with ${\bf a}_1=(1,0)$ and ${\bf a}_2=(1/2,\sqrt{3}/2)$ being the lattice vectors of the ordering pattern (see Fig.~\ref{fig:kagome1}). Existence of a flat band in the spin-wave dispersions implies one of the singular values of $\mathcal{R}({\bf k})$ is always 0 at any ${\bf k}$ i.e. a rank reduction of $\mathcal{R}({\bf k})$ by 1 which effectively makes it describable in terms of $\nu=\pm1$ block matrices as explained previously in Sect.~\ref{secfour}.  

The $q=0$ coplanar order has a $\tilde{\mathcal{C}}_2$ rotation symmetry (a rotation of 180$^\circ$) about each of the lattice sites in the unit cell which, in the momentum space, acts as $\mathcal{R}(\bf k)\rightarrow \mathcal{R}(-{\bf k})$. Since the complex conjugation $\mathcal{K}$ also does so, a combination of these two $\mathcal{T}\equiv\mathcal{C}_2\mathcal{K}$ is a symmetry of $\mathcal{R}({\bf k})$ where $\mathcal{T}$ is a antiunitary operator with $\mathcal{T}^2=1$. Consequently, expressed in a $\mathcal{T}$-invariant basis, $\mathcal{R}_{1,2}$ have real elements in the entire BZ. The classifying space of the SVD flattened matrices for $\mathcal{R}_{1}$ and $\mathcal{R}_{2}$ are $SO(4)$ and $SO(3)$ respectively, both having a  $\mathds{Z}_2$ topology as suggested by the table of $\nu\neq0$ systems. As a result, we observe Dirac strings in the BZ emanating from the $\Gamma$ point, and the topological invariant $\zeta$ (Eq.~\ref{topinv3}) for any closed loop surrounding the $\Gamma$ point is $-1$ [Fig.~\ref{fig:kagome2} (a) bottom]. The robust nature of the zero modes in form of such a flat band for the $q=0$ coplanar state has a topological origin typified by the $\mathds{Z}_2$ valued invariant $\zeta$.

\subsubsection{Inclusion of anisotropic scalar exchanges}

The flat band of zero modes persists even in anisotropic kagome Heisenberg models as long as the $q=0$ state has an analog of a flat sheet origami. There are examples of kagome materials of this kind~\cite{roychowdhury2017spin}. The simplest one is a $J_1-J_2-J_3$ type Heisenberg model for which the $q=0$ spin pattern has the same unit cell as the lattice [Fig.~\ref{fig:kagome1} (b)]. The spin Hamiltonian is given in Eq.~\ref{Hspinfrust} with $S_{\triangle\alpha}=a_1S_{1\alpha}+a_2S_{2\alpha}+a_3S_{3\alpha}$ in each unit cell and $J^{\triangle\alpha,\triangle'\beta}=J\delta^{\triangle\triangle'}\delta^{\alpha\beta}$ for all the interactions are between nn only and that  $J_{1}=Ja_{2}a_{3}$, $J_{2}=Ja_{3}a_{1}$, $J_{3}=Ja_{1}a_{2}$. The isotropic limit is simply given by $a_1=a_2=a_3=1$. When we vary the ratios $a_{2}/a_1$ and $a_{3}/a_1$ away from 1, we alter the strength of the anisotropic exchanges in the model. The spin-wave dispersions for this model is plotted in Fig.~\ref{fig:kagome2} (b) top with $a_1=1$, $a_2=2$, and $a_3$ determined by the constraint $a_1{\bf S}_1+a_2{\bf S}_2+a_3{\bf S}_3=0$ where ${\bf S}_1$, ${\bf S}_2$, and ${\bf S}_3$ are unit vectors forming an equilateral triangle. Remarkably the effects are only to change the locations of the Dirac strings while the value of $\zeta(=-1)$ remains invariant [Fig.~\ref{fig:kagome2} (b) bottom]. This explains the immunity of the flat band of zero modes against certain anisotropic scalar perturbations, thus, signifies the role of topology in rendering robustness to frustration as emphasized in this article.   

\subsubsection{Inclusion of spin-orbit coupling}

A further generalization of the constraint functions allows for various kinds of symmetric and antisymmetric spin-orbit exchanges to be incorporated into the spin Hamiltonian in Eq.~\ref{ham2}. However, given a generic spin model specified by a Hamiltonian $H_{\rm spin}$ including all such interactions may not be cast in terms of constraints like ${\bf S}_{\triangle}$, such that $H_{\rm spin}={\bf S}^2_{\triangle}+{\rm const.}$. Investigating this issue is beyond the reaches of the present work to bypass which we rather tweak the constraint functions first and then illustrate what kinds of interactions do they generate that preserve the flat band in the spin-wave dispersions. The modification is to multiply the scalars $a_{(1,2,3)}$ by orthogonal matrices $\mathds{M}_{(1,2,3)}$ and write the constraint functions as ${\bf S}_{\triangle} = a_1\mathds{M}_1\cdot{\bf S}_{1} + a_2\mathds{M}_2\cdot{\bf S}_{2} + a_3\mathds{M}_3\cdot{\bf S}_{3}$. Evidently, the trace of a term like $\mathds{M}_i^T\mathds{M}_j$ would lead to the anisotropic scalar exchanges, while the traceless symmetric part and the antisymmetric part of $\mathds{M}_i^T\mathds{M}_j$ would contribute to the symmetric spin-orbit exchanges and the Dzyaloshinskii-Moriya (DM) type interactions between the spins respectively.  

To this end, we distinguish between two different types of $\mathds{M}$ matrices for reasons to be clear soon. The first class of matrices add only anisotropic scalar exchanges and antisymmetric DM terms to the Hamiltonian and can be parametrized as
\begin{equation}
\mathds{M}_j = 
 \begin{pmatrix}
  \cos\theta_j & \sin\theta_j & 0 \\
  -\sin\theta_j & \cos\theta_j & 0 \\
  0 & 0 & 1 \\
 \end{pmatrix}
 \label{Mmat}
\end{equation}
implying the DM vector pointing along the $z$-axis and that the $SO(3)$ spin rotation symmetry of the Hamiltonian is broken down to $SO(2)$. Such perturbations retain the block-diagonal form of $\mathcal{R}$. To study their effects on the topology of $\mathcal{R}_{1,2}$, we consider $a_1=a_2=1$, $\mathds{M}_1=\mathds{I}$, and vary $\theta_2$ away from $0$, while the constraint ${\bf S}_{\triangle} = 0$ decides the values of $a_3$ and $\mathds{M}_3$. The top panel of Fig.~\ref{fig:kagome2} (c) shows the spin-wave dispersions for $\theta_2=\pi/5$. We observe that such variations only alter the locations of the Dirac strings, thus, preserve the topology of $\mathcal{R}$ as seen in the bottom panel of Fig.~\ref{fig:kagome2} (c).

The second class of matrices are taken as generic orthogonal matrices which add all sorts of interactions (anisotropic scalar exchanges, antisymmetric DM terms, and symmetric spin-orbit exchanges) to the Hamiltonian and break the $SO(3)$ spin rotation symmetry completely. We consider the following parametric form of such matrices
\begin{equation}
\mathds{M}_j = {\rm Exp}[\theta_j\bm{\omega}_j\cdot {\bf L}],
\label{so3brokenmat}
\end{equation}
where $\bm{\omega}_j$ is a unit vector specified as 
\begin{equation}
\bm{\omega}_j = (\cos\eta_j\sin\xi_j,\sin\eta_j\sin\xi_j,\cos\xi_j),
\label{so3brokenomega}
\end{equation}
and $L^a_{bc}=\epsilon_{abc}$ (the completely antisymmetric Levi-Civita tensor) are the generators of the $SO(3)$ group. Such types of perturbations mix the blocks $\mathcal{R}_{1}$ and $\mathcal{R}_{2}$, and in that case we must analyze the topology of a $SO(6)$ matrix corresponding to a $\nu=0$ system. Nevertheless, we observe Dirac strings that protect the flat band of zero modes even in the presence of all different kinds of spin exchanges. The top panel of Fig.~\ref{fig:kagome2} (d) shows the effects of such perturbations on the spin-wave dispersions for the parameters $a_1=a_2=1$, $\theta_1=\eta_1=\xi_1=0$, $\eta_2=\xi_2=1$, $\theta_2=\pi/10$ while $a_3$ and $\mathds{M}_3$ are decided by the constraint ${\bf S}_{\triangle}=0$ as before. However, the topological invariant $\zeta=1$ for this model calculated along any close contour around the $\Gamma$ point [Fig.~\ref{fig:kagome2} (d) bottom]. In summary, all these perturbations retain the flat band (although modify the frequencies of the non-flat bands) and its topology which evidences the robust nature of zero modes in certain classes of frustrated magnets.

\subsection{Other examples of $\nu\neq0$ frustrated systems}

Some of the other examples of $\nu\neq0$ frustrated systems include the pyrochlore magnets~\cite{reimers1992absence, moessner1998properties} and their projected versions onto two dimensions which are the checkerboard magnets (Heisenberg model on a checkerboard lattice~\cite{lieb1999ground, canals2002square}), both of which have been thoroughly studied in past for their fame of harboring exotic states of matter as a consequence of high frustration. The constraints in the spin Hamiltonian are that the total spin vanishes in each tetrahedra in the former and in each checkerboard in the latter. The MCM index for them is $\nu=2$ and $\nu=1$ respectively which envisage, following our tables, that the degeneracy of zero modes in these systems are also protected by a similar topology discussed above. 

\section{Conclusions}\label{secseven}

In conclusion, we explore a fundamental connection between magnetic frustration and topology, namely, how different forms of zero modes in a frustrated system can be topologically classified. The frustrated models of our concern share features with metamaterial Hamiltonians, and so, in uncovering their topological aspects, recent developments in the field of topological mechanics turn out to be extremely useful. Specifically, all the zero modes (zero to linear order) in a frustrated model/metamaterial can be explained in the framework of rigidity matrices (whose kernel contains the zero modes) $\mathcal{R}$ and the (linearized) Hamiltonian can be cast in a bilinear form in terms of $\mathcal{R}$. The key to decode the topology that protects the degeneracy of the zero modes in form of either isolated points (like Weyl points) or line nodes or surfaces (like flat bands) is to study the classifying spaces of these matrices in presence of various unitary and antiunitary symmetries of the problem. In this context, we present the striking result that even non-square rigidity matrices (i.e non-isostatic systems) with a non-zero Maxwell index $\nu$ exist in a non-trivial topological space. Thus our results introduce new classes of topological mechanical systems beyond the original Kane and Lubensky\cite{kane2014topological} isostatic class. 

To summarize our specific results, we present a classification of rigidity matrices guided by the ten-fold way of electronic band insulators and superconductors. This provides an explanation of zero modes in frustrated systems/metamaterials from topology. The class depends only on the absence or presence of the antiunitary time-reversal symmetry $\mathcal{T}$ in contrast to the ten-fold way that includes particle-hole symmetry and chiral symmetry in addition to $\mathcal{T}$ and is thus a three-fold way. However, unlike the ten-fold way which deals with Hamiltonian matrices, the key element in our discussion is the rigidity matrix which is non-Hermitian. To classify such non-Hermitian matrices we employ SVD flattening of rigidity matrices under the presence or absence of $\mathcal{T}$ (instead of spectral flattening of Hamiltonians) which lead us to the rich structures of Stiefel manifold in distinction to the Grassmannian manifold of the ten-fold way. We further study the different homotopy groups of the Stiefel manifolds which are endowed with intriguing topological structures revealing new topological invariants beyond those in the ten-fold classification table. Thus we expect new forms of zero modes will be found that are yet to be discovered in frustrated spin systems/metamaterials. We illustrate our claims by providing a number of emblematic examples of frustrate spin models that include the flat band in kagome Heisenberg systems and the $J_1-J_2$ Heisenberg model on a square lattice. We demonstrate how the physics of frustration in those non-isostatic $\nu \neq 0$ systems can be captured by real rigidity matrices and associated zero modes demanded by a vortex-like topological invariant. 

We believe these results are so general that this classification of rigidity matrices will elucidate the origin of frustration in the form of accidental degeneracy in a wide class of frustrated magnets by relating it to topological invariants that protect the robust nature of their zero modes. Perhaps the most promising application of these results is the explanation of accidental degeneracy found in the spin wave spectra of a magnetic insulator derived from neutron scattering data. Our example calculations suggest such spin waves arise from an ordering pattern which is characterized by a set of local constraints. These in turn create a rigidity matrix upon linearization and through it a set of topological invariants (of either the $\mathbb{Z}_2$, $\mathbb{Z}$ variety or more exotic  $\mathbb{Z}_{24}$, $\mathbb{Z}_{12}\times \mathbb{Z}_2$, etc. variety), whose changes demand the discovered accidental degeneracy. Such an explanation would then produce a prediction on how to control the degeneracy via perturbations which either keep or destroy the topological invariants. Finally, these predictions, beyond illuminating new properties of magnetic phases, would enable the search for exotic phases of matter that naturally arise from frustration such as spin ices and quantum spin liquids.

\section{Acknowledgements}

We thank Andreas W. W. Ludwig, D. Zeb Rocklin and Martin Zirnbauer for illuminating discussion. KR and MJL acknowledge supported in part by the National Science Foundation under Grant No. NSF PHY17-48958.

\bibliography{reference_ordered}

\end{document}